\documentclass[fleqn,usenatbib]{mnras}

\usepackage{newtxtext,newtxmath}
\usepackage{subfig}
\usepackage{comment}
\usepackage{longtable}
\usepackage{caption}
\usepackage{placeins}
\usepackage{threeparttable}
\newcommand{\be}{\begin{equation}}
\newcommand{\ee}{\end{equation}}
\newcommand{\bea}{\begin{eqnarray}}
\newcommand{\eea}{\end{eqnarray}}
\newcommand{\hunit}{$\rm{km \ s^{-1} \ Mpc^{-1}}$}
\newcommand{\lcdm}{$\Lambda$CDM}
\newcommand{\pcdm}{$\phi$CDM}
\usepackage{array}
\makeatletter
\newcommand{\thickhline}{%
    \noalign {\ifnum 0=`}\fi \hrule height 1pt
    \futurelet \reserved@a \@xhline
}
\newcolumntype{"}{@{\hskip\tabcolsep\vrule width 1pt\hskip\tabcolsep}}
\makeatother

\newcommand{\hiig}{HIIG}
\newcommand{\om}{$\Omega_{\rm m_0}$}

\usepackage[T1]{fontenc}

\DeclareRobustCommand{\VAN}[3]{#2}
\let\VANthebibliography\thebibliography
\def\thebibliography{\DeclareRobustCommand{\VAN}[3]{##3}\VANthebibliography}

\usepackage{graphicx}	
\usepackage{amsmath}	
\usepackage{amssymb}	

\title[Cosmological constraints from HIIG measurements]{Cosmological constraints from HII starburst galaxy apparent magnitude and other cosmological measurements}

\author[Cao et al.]{
Shulei Cao,$^{1}$\thanks{E-mail: shulei@phys.ksu.edu}
Joseph Ryan,$^{1}$\thanks{E-mail: jwryan@phys.ksu.edu}
Bharat Ratra$^{1}$\thanks{E-mail: ratra@phys.ksu.edu}
\\
$^{1}$Department of Physics, Kansas State University, 116 Cardwell Hall, Manhattan, KS 66502, USA
}

\date{Accepted XXX. Received YYY; in original form ZZZ}

\pubyear{2020}

\begin{document}
\label{firstpage}
\pagerange{\pageref{firstpage}--\pageref{lastpage}}
\maketitle

\begin{abstract}
We use HII starburst galaxy apparent magnitude measurements to constrain cosmological parameters in six cosmological models. A joint analysis of HII galaxy, quasar angular size, baryon acoustic oscillations peak length scale, and Hubble parameter measurements result in relatively model-independent and restrictive estimates of the current values of the non-relativistic matter density parameter \om\ and the Hubble constant $H_0$. These estimates favor a 2.0$\sigma$ to 3.4$\sigma$ (depending on cosmological model) lower $H_0$ than what is measured from the local expansion rate. The combined data are consistent with dark energy being a cosmological constant and with flat spatial hypersurfaces, but do not strongly rule out mild dark energy dynamics or slightly non-flat spatial geometries.
\end{abstract}




\section{Introduction} \label{sec:intro}

The accelerated expansion of the current universe is now well-established observationally and is usually credited to a dark energy whose origins remain murky (see e.g. \citealp{Ratra_Vogeley,Martin,Coley_Ellis}). The standard \lcdm\ model of cosmology \citep{peeb84} describes a universe with flat spatial hypersurfaces predominantly filled with dark energy in the form of a cosmological constant $\Lambda$ and cold dark matter (CDM) together comprising $\sim95$\% of the total energy budget. While spatially-flat \lcdm\ is mostly consistent with cosmological observations (see e.g. \citealp{Alam_et_al_2017,Farooq_Ranjeet_Crandall_Ratra_2017,scolnic_et_al_2018,planck2018b}), there are indications of some (mild) discrepances between standard \lcdm\ model predictions and cosmological measurements. In addition, the quality and quantity of cosmological data continue to grow, making it possible to consider and constrain additional cosmological parameters beyond those that characterize the standard \lcdm\ model.

Given the uncertainty surrounding the origin of the cosmological constant, many workers have investigated the possibility that the cosmological ``constant'' is not really constant, but rather evolves in time, either by positing an equation of state parameter $w \neq -1$ (thereby introducing a redshift dependence into the dark energy density) or by replacing the constant $\Lambda$ in the Einstein-Hilbert action with a dynamical scalar field $\phi$ (\citealp{peebrat88, ratpeeb88}). Non-flat spatial geometry also introduces a time-dependent source term in the Friedmann equations. In this paper we study the standard spatially-flat \lcdm\ model as well as dynamical dark energy and spatially non-flat extensions of this model.

One major goal of this paper is to use measurements of the redshift, apparent luminosity, and gas velocity dispersion of HII starburst galaxies to constrain cosmological parameters.\footnote{For early attempts see \cite{Siegel_2005}, \cite{Plionis_2009,Plionis_2010,Plionis_2011} and \cite{Mania_2012}. For more recent discussions see \cite{Chavez_2016}, \cite{Wei_2016}, \cite{Yennapureddy_2017}, \cite{Zheng_2019}, \cite{ruan_etal_2019}, \cite{G-M_2019}, \cite{Wan_2019}, and \cite{Wu_2020}.}
An HII starburst galaxy (hereinafter ``HIIG'') is one that contains a large HII region, an emission nebula sourced by the UV radiation from an O- or B-type star. There is a correlation between the measured luminosity ($L$) and the inferred velocity dispersion ($\sigma$) of the ionized gases within these HIIG, referred to as the $L$-$\sigma$ relation (see Sec. \ref{sec:data}) which has been shown to be a useful cosmological tracer (see \citealp{Melnick_2000,Siegel_2005,Plionis_2011,Chavez_2012,Chavez_2014,Chavez_2016,Terlevich_2015,G-M_2019}, and references therein). This relation has been used to constrain the Hubble constant $H_0$ (\citealp{Chavez_2012,FernandezArenas}), and it can also be used to put constraints on the dark energy equation of state parameter $w$ (\citealp{Terlevich_2015,Chavez_2016,G-M_2019}).

HIIG data reach to redshift $z\sim2.4$, a little beyond that of the highest redshift baryon acoustic oscillation (BAO) data which reach to $z\sim2.3$. HIIG data are among a handful of cosmological observations that probe the largely unexplored part of redshift space from $z\sim2$ to $z\sim1100$. Other data that probe this region include quasar angular size measurements that reach to $z\sim2.7$ (\citealp{gurvits_kellermann_frey_1999,chen_ratra_2002,Cao_et_al2017b,Ryan_2}, and references therein), quasar flux measurements that reach to  $z\sim5$ (\citealp{RisalitiandLusso_2015,RisalitiandLusso_2019,Yang_2019,KhadkaandRatra_2020,Khadka_2020b,Zheng_2020}, and references therein), and gamma ray burst data that reach to $z\sim8$ (\citealp{Lamb_2000,samushia_ratra_2010,Demianski_2019}, and references therein). In this paper we also use quasar angular size measurements (hereinafter ``QSO'') to constrain cosmological model parameters.

While HIIG and QSO data probe the largely unexplored $z\sim2.3$--2.7 part of the universe, current \hiig\ and QSO measurements provide relatively weaker constraints on cosmological parameters than those provided by more widely used measurements, such as BAO peak length scale observations or Hubble parameter (hereinafter ``$H(z)$'') observations (with these latter data being at lower redshift but of better quality than \hiig\ or QSO data). However, we find that the HIIG and QSO constraints are consistent with those that follow from BAO and $H(z)$ data, and so we use all four sets of data together to constrain cosmological parameters. We find that the HIIG and QSO data tighten parameter constraints relative to the $H(z)$ + BAO only case.

Using six different cosmological models to constrain cosmological parameters allows us to determine which of our results are less model-dependent. In all models, the HIIG data favor those parts of cosmological parameter space for which the current cosmological expansion is accelerating.\footnote{This result could weaken, however, as the \hiig\ data constraint contours could broaden when \hiig\ data systematic uncertainties are taken into account. We do not incorporate any \hiig\ systematic uncertainties into our analysis; see below.} The joint analysis of the HIIG, QSO, BAO and $H(z)$ data results in relatively model-independent and fairly tight determination of the Hubble constant $H_0$ and the current non-relativistic matter density parameter \om.\footnote{The BAO and $H(z)$ data play a more significant role than do the HIIG and QSO data in setting these and other limits, but the \hiig\ and QSO data tighten the BAO + $H(z)$ constraints. We note, however, that the $H(z)$ and QSO data, by themselves, give lower central values of $H_0$ but with larger error bars. Also, because we calibrate the distance scale of the BAO measurements listed in Table \ref{tab:BAO} via the sound horizon scale at the drag epoch ($r_s$, about which see below), a quantity that depends on early-Universe physics, we would expect these measurements to push the best-fit values $H_0$ lower when they are combined with late-Universe measurements like \hiig\ (whose distance scale is not set by the physics of the early Universe).} Depending on the model, \om\ ranges from a low of $0.309^{+0.015}_{-0.014}$ to a high of $0.319 \pm 0.013$, being consistent with most other estimates of this parameter (unless indicated otherwise, uncertainties given in this paper are $\pm 1\sigma$). The best-fit values of $H_0$, ranging from $68.18^{+0.97}_{-0.75}$ \hunit to $69.90 \pm 1.48$ \hunit, are, from the quadrature sum of the error bars, 2.01$\sigma$ to 3.40$\sigma$ lower than the local $H_0 = 74.03 \pm 1.42$ \hunit measurement of \cite{riess_etal_2019} and only 0.06$\sigma$ to 0.60$\sigma$ higher than the median statistics $H_0 = 68 \pm 2.8$ \hunit estimate of \cite{chenratmed}. These combined measurements are consistent with the spatially-flat \lcdm\ model, but also do not strongly disallow some mild dark energy dynamics, as well as a little non-zero spatial curvature energy density.

This paper is organized as follows. In Sec. \ref{sec:data} we introduce the data we use. Section \ref{sec:model} describes the models we analyze, with a description of our analysis method in Sec. \ref{sec:analysis}. Our results are in Sec. \ref{sec:results}, and we provide our conclusions in Sec. \ref{sec:conclusion}. 

\section{Data}
\label{sec:data}
We use a combination of $H(z)$, BAO, QSO, and HIIG data to obtain constraints on our cosmological models. The $H(z)$ data, spanning the redshift range $0.070 \leq z \leq 1.965$, are identical to the $H(z)$ data used in \cite{Ryan_1, Ryan_2} and compiled in Table 2 of \cite{Ryan_1}; see that paper for description. The QSO data compiled by \cite{Cao_et_al2017b} (listed in Table 1 of that paper) and spanning the redshift range $0.462 \leq z \leq 2.73$, are identical to that used in \cite{Ryan_2}; see these papers for descriptions. Our BAO data (see Table \ref{tab:BAO}) have been updated relative to \cite{Ryan_2} and span the redshift range $0.38 \leq z \leq 2.34$. Our HIIG data are new, comprising 107 low redshift ($0.0088 \leq z \leq 0.16417$) HIIG measurements, used in \cite{Chavez_2014}, and 46 high redshift ($0.636427 \leq z \leq 2.42935$) HIIG measurements, used in \cite{G-M_2019}.\footnote{15 from \cite{G-M_2019}, 25 from \cite{Erb_2006}, \cite{Masters_2014}, and \cite{Maseda_2014}, and 6 from \cite{Terlevich_2015}.} These extinction-corrected measurements (see below for a discussion of extinction correction) were very kindly provided to us by Ana Luisa Gonz\'{a}lez-Mor\'{a}n (private communications, 2019 and 2020).

In order to use BAO measurements to constrain cosmological model parameters, knowledge of the sound horizon scale at the drag epoch ($r_s$) is required. We compute this scale more accurately than in \cite{Ryan_2} by using the approximate formula \citep{PhysRevD.92.123516}
\be
\label{eq:sh}
    r_s=\frac{55.154\exp{[-72.3(\Omega_{\rm \nu_0}h^2+0.0006)^2]}}{(\Omega_{\rm b_0}h^2)^{0.12807}(\Omega_{\rm cb_0}h^2)^{0.25351}}.
\ee
Here $\Omega_{\rm cb_0} = \Omega_{\rm c_0} + \Omega_{\rm b_0} = \Omega_{\rm m_0} - \Omega_{\rm \nu_0}$ with $\Omega_{\rm cb_0}$, $\Omega_{\rm c_0}$, $\Omega_{\rm b_0}$, and $\Omega_{\nu_0} = 0.0014$ (following \citealp{Carter_2018}) being the current values of the CDM + baryonic matter, CDM, baryonic matter, and neutrino energy density parameters, respectively, and the Hubble constant $H_0 = 100\ h$ \hunit. Here and in what follows, a subscript of `0' on a given quantity denotes the current value of that quantity. Additionally, $\Omega_{\rm b_0}h^2$ is slightly model-dependent; the values of this parameter that we use in this paper are the same as those originally computed in \cite{park_ratra_2018, park_ratra_2019a, park_ratra_2019c} and listed in Table 2 of \cite{Ryan_2}.

\begin{table}
\centering
\begin{threeparttable}
\caption{BAO data.}\label{tab:BAO}
\setlength{\tabcolsep}{1.5mm}{
\begin{tabular}{lccc}
\thickhline
$z$ & Measurement\tnote{a} & Value & Ref.\\
\hline
$0.38$ & $D_M\left(r_{s,{\rm fid}}/r_s\right)$ & 1512.39 & \cite{Alam_et_al_2017}\tnote{b}\\
\hline
$0.38$ & $H(z)\left(r_s/r_{s,{\rm fid}}\right)$ & 81.2087 & \cite{Alam_et_al_2017}\tnote{b}\\
\hline
$0.51$ & $D_M\left(r_{s,{\rm fid}}/r_s\right)$ & 1975.22 & \cite{Alam_et_al_2017}\tnote{b}\\
\hline
$0.51$ & $H(z)\left(r_s/r_{s,{\rm fid}}\right)$ & 90.9029 & \cite{Alam_et_al_2017}\tnote{b}\\
\hline
$0.61$ & $D_M\left(r_{s,{\rm fid}}/r_s\right)$ & 2306.68 & \cite{Alam_et_al_2017}\tnote{b}\\
\hline
$0.61$ & $H(z)\left(r_s/r_{s,{\rm fid}}\right)$ & 98.9647 & \cite{Alam_et_al_2017}\tnote{b}\\
\hline
$0.122$ & $D_V\left(r_{s,{\rm fid}}/r_s\right)$ & $539\pm17$ & \cite{Carter_2018}\\
\hline
$0.81$ & $D_A/r_s$ & $10.75\pm0.43$ & \cite{DES_2019b}\\
\hline
$1.52$ & $D_V\left(r_{s,{\rm fid}}/r_s\right)$ & $3843\pm147$ & \cite{3}\\
\hline
$2.34$ & $D_H/r_s$ & 8.86 & \cite{Agathe}\tnote{c}\\
\hline
$2.34$ & $D_M/r_s$ & 37.41 & \cite{Agathe}\tnote{c}\\
\thickhline
\end{tabular}}
\begin{tablenotes}
\item[a] $D_M \left(r_{s,{\rm fid}}/r_s\right)$, $D_V \left(r_{s,{\rm fid}}/r_s\right)$, $r_s$, and $r_{s, {\rm fid}}$ have units of Mpc, while $H(z)\left(r_s/r_{s,{\rm fid}}\right)$ has units of \hunit, and $D_A/r_s$ is dimensionless.
\item[b] The six measurements from \cite{Alam_et_al_2017} are correlated; see eq. (20) of \cite{Ryan_2} for their correlation matrix.
\item[c] The two measurements from \cite{Agathe} are correlated; see eq. \eqref{CovM} below for their correlation matrix.
\end{tablenotes}
\end{threeparttable}
\end{table}

As mentioned in Sec. \ref{sec:intro}, HIIG can be used as cosmological probes because they exhibit a tight correlation between the observed luminosity ($L$) of their Balmer emission lines and the velocity dispersion ($\sigma$) of their ionized gas (as measured from the widths of the emission lines). That correlation can be expressed in the form
\begin{equation}
\label{eq:logL}
    \log L = \beta \log \sigma + \gamma,
\end{equation}
where $\gamma$ and $\beta$ are the intercept and slope, respectively, and $\log = \log_{10}$ here and in what follows. In order to determine the values of $\beta$ and $\gamma$, it is necessary to establish the extent to which light from an HIIG is extinguished as it propagates through space. A correction must be made to the observed flux so as to account for the effect of this extinction. As mentioned above, the data we received from Ana Luisa Gonz\'{a}lez-Mor\'{a}n have been corrected for extinction. In \cite{G-M_2019}, the authors used the \cite{Gordon_2003} extinction law, and in so doing found
\begin{equation}
    \label{eq:Gordon_beta}
    \beta = 5.022 \pm 0.058,
\end{equation}
and
\begin{equation}
    \label{eq:Gordon_gamma}
    \gamma = 33.268 \pm 0.083,
\end{equation}
respectively. These are the values of $\beta$ and $\gamma$ that we use in the $L$-$\sigma$ relation, eq. \eqref{eq:logL}.

Once the luminosity of an HIIG has been established through eq. \eqref{eq:logL}, this luminosity can be used, in conjunction with a measurement of the flux ($f$) emitted by the HIIG, to determine the distance modulus of the HIIG via
\begin{equation}
    \mu_{\rm obs} = 2.5\log L - 2.5\log f - 100.2
\end{equation}
(see e.g. \citealp{Terlevich_2015}, \citealp{G-M_2019}, and references therein).\footnote{For each HIIG in our sample we have the measured values and uncertainties of $\log f$\!, $\log \sigma$\!, and $z$.} This quantity can then be compared to the value of the distance modulus predicted within a given cosmological model
\be
\label{eq:mu_th}
    \mu_{\rm th}\left(\textbf{p}, z\right) = 5\log D_{\rm L}\left(\textbf{p}, z\right) + 25,
\ee
where the luminosity distance $D_L(\textbf{p}, z)$ is related to the transverse comoving distance $D_M(\textbf{p}, z)$ and the angular size distance $D_A(\textbf{p}, z)$ through $D_L(\textbf{p}, z)=(1+z)D_M(\textbf{p}, z)=(1+z)^2D_A(\textbf{p}, z)$. These are functions of the redshift $z$ and the parameters $\textbf{p}$ of the model in question, and
\be
\label{eq:DL}
\resizebox{0.44\textwidth}{!}{%
  $D_M(\textbf{p}, z) = 
    \begin{cases}
    \vspace{1mm}
    D_C(\textbf{p}, z) & \text{if}\ \Omega_{\rm k_0} = 0,\\
    \vspace{1mm}
    \frac{c}{H_0\sqrt{\Omega_{\rm k_0}}}\sinh\left[\sqrt{\Omega_{\rm k_0}}H_0D_C(\textbf{p}, z)/c\right] & \text{if}\ \Omega_{\rm k_0} > 0, \\
    \vspace{1mm}
    \frac{c}{H_0\sqrt{|\Omega_{\rm k_0}|}}\sin\left[\sqrt{|\Omega_{\rm k_0}|}H_0D_C(\textbf{p}, z)/c\right] & \text{if}\ \Omega_{\rm k_0} < 0.
    \end{cases}$%
    }
\ee
In the preceding equation,
\be
\label{eq:DC}
    D_C(\textbf{p}, z) \equiv c\int^z_0 \frac{dz'}{H(\textbf{p}, z')},
\ee
$\Omega_{\rm k_0}$ is the current value of the spatial curvature energy density parameter, and $c$ is the speed of light (\citealp{Hogg}).

As the precision of cosmological observations has grown over the last few years, a tension between measurements of the Hubble constant made with early-Universe probes and measurements made with late-Universe probes has revealed itself (for a review, see \citealp{Riess_faster}). Whether a given cosmological observation supports a lower value of $H_0$ (i.e. one that is closer to the early-Universe \textit{Planck} measurement) or a higher value of $H_0$ (i.e. one that is closer to the late-Universe value measured by \citealp{riess_etal_2019}) may depend on whether the distance scale associated with this observation has been set by early- or late-Universe physics. It is therefore important to know what distance scale cosmological observations have been calibrated to, so that the extent to which measurements of $H_0$ are pushed higher or lower by these different distance calibrations can be clearly identified.

The $H_0$ values we measure from the combined $H(z)$, BAO, QSO, and \hiig\ data are based on a combination of both early- and late-Universe distance calibrations. As mentioned above, the distance scale of our BAO measurements is set by the size of the sound horizon at the drag epoch $r_s$. The sound horizon, in turn, depends on $\Omega_{\rm b_0}h^2$, which was computed by \cite{park_ratra_2018, park_ratra_2019a, park_ratra_2019c} using early-Universe data. Our \hiig\ measurements, on the other hand, have been calibrated using cosmological model independent distance ladder measurements of the distances to nearby giant HII regions (see \citealp{G-M_2019} and references therein), so these data qualify as late-Universe probes. The distance scale of our QSO measurements is set by the intrinsic linear size ($l_m$) of the QSOs themselves, which is a late-Universe measurement (see \citealp{Cao_et_al2017b}). Finally, our $H(z)$ data depend on late-Universe astrophysics through the modeling of the star formation histories of the galaxies whose ages are measured to obtain the Hubble parameter (although the differences between different models are not thought to have a significant effect on measurements of $H(z)$ from these galaxies; see \citealp{Setting_stage_1, Setting_stage_2}).

\section{Cosmological models}
\label{sec:model}

The redshift $z$ is related to the scale factor $a$ as
$1 + z\equiv {a_0}/{a}$
and the Hubble parameter is $H\equiv{\dot{a}}/{a}$,
where the overdot denotes the time derivative. In this paper we consider three pairs of flat and non-flat cosmological models.\footnote{Observational constraints on non-flat models are discussed in \cite{Farooq_Mania_Ratra_2015}, \cite{Chen_et_al_2016}, \cite{yu_wang_2016}, \cite{rana_jain_mahajan_mukherjee_2017}, \cite{ooba_etal_2018a, ooba_etal_2018b, ooba_etal_2018c}, \cite{yu_etal_2018}, \cite{park_ratra_2018, park_ratra_2019a, park_ratra_2019b, park_ratra_2019c, park_ratra_2020}, \cite{wei_2018}, \cite{DES_2019}, \cite{ruan_etal_2019}, \cite{coley_2019}, \cite{jesus_etal_2019}, \cite{handley_2019}, \cite{wang_etal_2019}, \cite{zhai_etal_2019}, \cite{li_etal_2020}, \cite{geng_etal_2020}, \cite{kumar_etal_2020}, \cite{efstathiou_gratton_2020}, \cite{divalentino_etal_2020}, \cite{gao_etal_2020}, and references therein.} The data we use are at $z \leq 2.73$ and in what follows we ignore the insignificant contribution that radiation makes to the late-time cosmological energy budget.

In the \lcdm\ model, the Hubble parameter is
\be\label{Hz1}
H(z)=H_0\sqrt{\Omega_{\rm m_0}(1+z)^3+\Omega_{\rm k_0}(1+z)^2+\Omega_{\Lambda}},
\ee
where $\Omega_{\Lambda}$ is the (constant) dark energy density parameter. In the flat \lcdm\ model the parameters to be constrained are conventionally chosen to be $H_0$ and $\Omega_{\rm m_0}$. In this model $\Omega_{\rm k_0}=0$, which implies $\Omega_{\Lambda}=1-\Omega_{\rm m_0}$. In the non-flat \lcdm\ model the parameters to be constrained are $H_0$, $\Omega_{\rm m_0}$, and $\Omega_{\Lambda}$, and the curvature energy density parameter is a derived quantity, being related to the non-relativistic matter and dark energy density parameters through $\Omega_{\rm k_0}=1-\Omega_{\rm m_0}-\Omega_{\Lambda}$.

In the XCDM parametrization, dark energy is modeled as an ideal, spatially homogeneous X-fluid with equation of state $w_{\rm X}=p_{\rm X}/\rho_{\rm X}$, where $p_{\rm X}$ and $\rho_{\rm X}$ are the X-fluid's pressure and energy density, respectively.\footnote{It should be noted, however, that the XCDM parametrization cannot sensibly describe the evolution of spatial inhomogeneities, and is therefore, unlike the \lcdm\ and \pcdm\ models, physically incomplete. It is possible to extend this parametrization by allowing for an additional free parameter $c^2_{s, {\rm X}} = dp_{\rm X}/d\rho_{\rm X}$ and requiring that $c^2_{s, {\rm X}} > 0$.} In the XCDM parametrization, the Hubble parameter is

\be\label{Hz2}
H(z)=H_0\sqrt{\Omega_{\rm m_0}(1+z)^3+\Omega_{\rm k_0}(1+z)^2+\Omega_{\rm X_0}(1+z)^{3(1+w_{\rm X})}},
\ee
where $\Omega_{\rm X_0}$ is the present value of the X-fluid energy density parameter. From this equation, it can be seen that when $w_{\rm X} = -1$ XCDM reduces to \lcdm. In the non-flat case the model parameters to be constrained are $H_0$, $\Omega_{\rm m_0}$, $\Omega_{\rm k_0}$, and $w_{\rm X}$, with $\Omega_{\rm X_0}=1-\Omega_{\rm m_0}-\Omega_{\rm k_0}$ as a derived parameter (we do not report constraints on its value in this paper). In the spatially-flat case the parameters to be constrained are $H_0$, $\Omega_{\rm m_0}$, and $w_{\rm X}$, with $\Omega_{\rm X_0}=1-\Omega_{\rm m_0}$.

In the flat and non-flat $\phi$CDM models, dark energy is modeled as a dynamical scalar field $\phi$, with a potential energy density given by

\be\label{PE}
V(\phi)=\frac{1}{2}\kappa m_p^2\phi^{-\alpha},
\ee
\\
where $m_p$ is the Planck mass, $\alpha$ is a non-negative scalar, and 
\be\label{kp}
\kappa=\frac{8}{3m_p^2}\bigg(\frac{\alpha+4}{\alpha+2}\bigg)
\bigg[\frac{2}{3}\alpha(\alpha+2)\bigg]^{\alpha/2}
\ee
(\citealp{peebrat88, ratpeeb88, pavlov13}).\footnote{Observational constraints on the \pcdm\ model are discussed in \cite{Chen_Ratra_2004}, \cite{samushia_et_al_2007}, \cite{yashar_et_al_2009}, \cite{Samushia_2010}, \cite{chen_ratra_2011b}, \cite{Campanelli_etal_2012}, \cite{Farooq_Ratra_2013}, \cite{farooq_crandall_ratra_2013}, \cite{Avsajanishvili_2015}, \cite{Sola_etal_2017}, \cite{zhai_blanton_slosar_tinker_2017}, \cite{sangwan_tripathi_jassal_2018}, \cite{Sola_perez_gomez_2018,sola_gomez_perez_2019}, \cite{ooba_etal_2019}, \cite{singh_etal_2019}, and references therein.} Note that when $\alpha=0$ the $\phi$CDM model reduces to the \lcdm\ model. In the spatially homogeneous approximation, valid for the cosmological tests we consider in this paper, the dynamics of the scalar field is governed by two coupled non-linear ordinary differential equations, the first being the scalar field equation of motion
\be\label{em}
\ddot{\phi}+3\bigg(\frac{\dot{a}}{a}\bigg)\dot{\phi}-\frac{1}{2}\alpha\kappa m_p^2\phi^{-\alpha-1}=0,
\ee
and the second being the Friedman equation
\be\label{fe}
\bigg(\frac{\dot{a}}{a}\bigg)^2=\frac{8\pi}{3m_p^2}(\rho_{\rm m}+\rho_{\phi})-\frac{k}{a^2}.
\ee
In eq. \eqref{fe}, ${-k}/{a^2}$ is the spatial curvature term (with $k = 0$, $-1$, $+1$ corresponding to $\Omega_{\rm k_0} = 0$, $>0$, $<0$, respectively), and $\rho_{\rm m}$ and $\rho_{\phi}$ are the non-relativistic matter and scalar field energy densities, respectively, where
\be\label{rp}
\rho_{\phi}=\frac{m_p^2}{32\pi}\bigg(\dot{\phi}^2+\kappa m_p^2\phi^{-\alpha}\bigg).
\ee
It follows that the Hubble parameter in $\phi$CDM is
\be\label{Hz3}
H(z)=H_0\sqrt{\Omega_{\rm m_0}(1+z)^3+\Omega_{\rm k_0}(1+z)^2+\Omega_{\phi}(z,\alpha)},
\ee
where the scalar field energy density parameter
\be\label{op}
\Omega_{\phi}(z,\alpha)=\frac{1}{12H_0^2}\bigg(\dot{\phi}^2+\kappa m_p^2\phi^{-\alpha}\bigg),
\ee
as can be determined from eqs. \eqref{em} and \eqref{fe}. For non-flat $\phi$CDM the parameters to be constrained are $\alpha$, $H_0$, $\Omega_{\rm m_0}$, and $\Omega_{\rm k_0}$ and for flat \pcdm\ the parameters to be constrained are $\alpha$, $H_0$, and $\Omega_{\rm m_0}$.

\section{Data Analysis Methodology}
\label{sec:analysis}

We perform a Markov chain Monte Carlo (MCMC) analysis with the Python module emcee \citep{2013PASP..125..306F} and maximize the likelihood function, $\mathcal{L}$, 
to determine the best-fit values of the parameters $\textbf{p}$ of the models. We use flat priors for all parameters $\textbf{p}$. For all models, the priors on $\Omega_{\rm m_0}$ and $h$ are non-zero over the ranges $0.1 \leq \Omega_{\rm m_0} \leq 0.7$ and $0.50 \leq h \leq 0.85$. In the non-flat \lcdm\ model the $\Omega_{\Lambda}$ prior is non-zero over $0.2 \leq \Omega_{\Lambda} \leq 1$. In the flat and non-flat XCDM parametrizations the prior range on $w_{\rm X}$ is $-2 \leq w_{\rm X} \leq 0$, and the prior range on $\Omega_{\rm k_0}$ in the non-flat XCDM parametrization is $-0.7 \leq \Omega_{\rm k_0} \leq 0.7$. In the flat and non-flat \pcdm\ models the prior range on $\alpha$ is $0.01 \leq \alpha \leq 3$ and $0.01 \leq \alpha \leq 5$, respectively, and the prior range on $\Omega_{\rm k_0}$ is also $-0.7 \leq \Omega_{\rm k_0} \leq 0.7$.

For HIIG, the likelihood function is
\be
\label{eq:LH1}
    \mathcal{L}_{\rm HIIG}= e^{-\chi^2_{\rm HIIG}/2},
\ee
where
\be
\label{eq:chi2_HIIG}
    \chi^2_{\rm HIIG}(\textbf{p}) = \sum^{153}_{i = 1} \frac{[\mu_{\rm th}(\textbf{p}, z_i) - \mu_{\rm obs}(z_i)]^2}{\epsilon_i^2},
\ee
and $\epsilon_i$ is the uncertainty of the $i_{\rm th}$ measurement. Following \cite{G-M_2019}, $\epsilon$ has the form
\be
\label{eq:err_HIIG}
    \epsilon=\sqrt{\epsilon^2_{\rm stat}+\epsilon^2_{\rm sys}},
\ee
where the statistical uncertainties are
\be
\label{eq:stat_err_HIIG}
    \epsilon^2_{\rm stat}=6.25\left[\epsilon^2_{\log f}+\beta^2\epsilon^2_{\log\sigma}+\epsilon^2_{\beta}(\log\sigma)^2+\epsilon^2_{\gamma}\right]+\left(\frac{\partial{\mu_{\rm th}}}{\partial{z}}\right)^2\epsilon^2_{z}.
\ee
Following \cite{G-M_2019} we do not account for systematic uncertainties in our analysis, so the uncertainty on the HIIG measurements consists entirely of the statistical uncertainty (so that $\epsilon = \epsilon_{\rm stat}$).\footnote{A systematic error budget for \hiig\ data is available in the literature, however; see \cite{Chavez_2016}.} The reader should also note here that although the theoretical statistical uncertainty depends our cosmological model parameters (through the theoretical distance modulus $\mu_{\rm th} = \mu_{\rm th}\left(\textbf{p}, z\right)$), the effect of this model-dependence on the parameter constraints is negligible for the current data.\footnote{In contrast to our definition of $\chi^2$ in eq. \eqref{eq:chi2_HIIG}, \cite{G-M_2019} defined an $H_0$-independent $\chi^2$ function in their eq. (27) and weighted this $\chi^2$ function by $1/\epsilon^2_{\rm stat}$ (where $\epsilon^2_{\rm stat}$ is given by their eq. (15)) which we do not do. This procedure is discussed in the literature \citep{Melnick_2017,FernandezArenas}, and when we use it we find that it leads to a reduced $\chi^2$ identical to that given in \cite{G-M_2019} (being less than 2 but greater than 1) without having a noticeable effect on the shapes or peak locations of our posterior likelihoods (hence providing very similar best-fit values and error bars of the cosmological model parameters). As discussed below, with our $\chi^2$ definition we find reduced $\chi^2$ values $\sim2.75$. \cite{G-M_2019} note that an accounting of systematic uncertainties could decrease the reduced $\chi^2$ values towards unity.\label{fn5}}

For $H(z)$, the likelihood function is
\be
\label{eq:LH2}
    \mathcal{L}_{\rm H}= e^{-\chi^2_{\rm H}/2},
\ee
where
\be
\label{eq:chi2_Hz}
    \chi^2_{\rm H}(\textbf{p}) = \sum^{31}_{i = 1} \frac{[H_{\rm th}(\textbf{p}, z_i) - H_{\rm obs}(z_i)]^2}{\epsilon_i^2},
\ee
and $\epsilon_i$ is the uncertainty of $H_{\rm obs}(z_i)$.

For the BAO data, the likelihood function is
\be
\label{eq:LH3}
    \mathcal{L}_{\rm BAO}= e^{-\chi^2_{\rm BAO}/2},
\ee
and for the uncorrelated BAO data (lines 7-9 in Table \ref{tab:BAO}) the $\chi^2$ function takes the form
\be
\label{eq:chi2_BAO1}
    \chi^2_{\rm BAO}(\textbf{p}) = \sum^{3}_{i = 1} \frac{[A_{\rm th}(\textbf{p}, z_i) - A_{\rm obs}(z_i)]^2}{\epsilon_i^2},
\ee
where $A_{\rm th}$ and $A_{\rm obs}$ are, respectively, the theoretical and observational quantities as listed in Table \ref{tab:BAO}, and $\epsilon_{i}$ corresponds to the uncertainty of $A_{\rm obs}(z_i)$. For the correlated BAO data, the $\chi^2$ function takes the form
\be
\label{eq:chi2_BAO2}
    \chi^2_{\rm BAO}(\textbf{p}) = [A_{\rm th}(\textbf{p}) - A_{\rm obs}(z_i)]^T\textbf{C}^{-1}[A_{\rm th}(\textbf{p}) - A_{\rm obs}(z_i)],
\ee
where superscripts $T$ and $-1$ denote the transpose and inverse of the matrices, respectively. The covariance matrix $\textbf{C}$ for the BAO data, taken from \cite{Alam_et_al_2017}, is given in eq. (20) of \cite{Ryan_2}, while for the BAO data from \cite{Agathe},
\be
\label{CovM}
    \textbf{C}=
    \begin{bmatrix}
    0.0841 & -0.183396 \\
    -0.183396 & 3.4596
    \end{bmatrix}.
\ee

For QSO, the likelihood function is
\be
\label{eq:LH4}
    \mathcal{L}_{\rm QSO}= e^{-\chi^2_{\rm QSO}/2},
\ee
and the $\chi^2$ function takes the form
\be
\label{eq:chi2_QSO}
    \chi^2_{\rm QSO}(\textbf{p}) = \sum^{120}_{i = 1} \left[\frac{\theta_{\rm th}(\textbf{p}, z_i) - \theta_{\rm obs}(z_i)}{\epsilon_i+0.1\theta_{\rm obs}(z_i)}\right]^2,
\ee
where $\theta_{\rm th}(\textbf{p}, z_i)$ and $\theta_{\rm obs}(z_i)$ are theoretical and observed values of the angular size at redshift $z_i$, respectively, and $\epsilon_i$ is the uncertainty of $\theta_{\rm obs}(z_i)$ (see \citealp{Ryan_2} for more details).

For the joint analysis of these data, the total likelihood function is obtained by multiplying the individual likelihood functions (that is, eqs. \eqref{eq:LH1}, \eqref{eq:LH2}, \eqref{eq:LH3}, and \eqref{eq:LH4}) together in various combinations. For example, for $H(z)$, BAO, and QSO data, we have
\be
\label{LH}
    \mathcal{L}=\mathcal{L}_{\rm H}\mathcal{L}_{\rm BAO}\mathcal{L}_{\rm QSO}.
\ee

We also use the Akaike Information
Criterion ($AIC$) and Bayesian Information Criterion ($BIC$) to compare the goodness of fit of models with different numbers of parameters, where
\be
\label{AIC}
    AIC=-2\ln \mathcal{L}_{\rm max} + 2n\equiv\chi^2_{\rm min}+2n,
\ee
and
\be
\label{BIC}
    BIC=-2\ln \mathcal{L}_{\rm max} + n\ln N\equiv\chi^2_{\rm min}+n\ln N.
\ee
In these two equations, $\mathcal{L}_{\rm max}$ refers to the maximum value of the given likelihood function, $\chi^2_{\rm min}$ refers to the corresponding minimum $\chi^2$ value, $n$ is the number of parameters of the given model, and $N$ is the number of data points (for example for HIIG we have $N=153$, etc.).

\section{Results}
\label{sec:results}
\subsection{HIIG constraints}
\label{subsec:HIIG}

We present the posterior one-dimensional (1D) probability distributions and two-dimensional (2D) confidence regions of the cosmological parameters for the six flat and non-flat models in Figs. \ref{fig01}--\ref{fig06}, in gray. The unmarginalized best-fit parameter values are listed in Table \ref{tab:BFP}, along with the corresponding $\chi^2$, $AIC$, $BIC$, and degrees of freedom $\nu$ (where $\nu \equiv N - n$). The marginalized best-fit parameter values and uncertainties ($\pm 1\sigma$ error bars or $2\sigma$ limits) are given in Table \ref{tab:1d_BFP}.\footnote{We plot these figures by using the Python package GetDist \citep{Lewis_2019}, which we also use to compute the central values (posterior means) and uncertainties of the cosmological parameters listed in Table \ref{tab:1d_BFP}.}

From the fit to the HIIG data, we see that most of the probability lies in the part of the parameter space corresponding to currently-accelerating cosmological expansion (see the gray contours in Figs. \ref{fig01}--\ref{fig06}). This means that the HIIG data favor currently-accelerating cosmological expansion,\footnote{Although a full accounting of the systematic uncertainties in the \hiig\ data could weaken this conclusion.} in agreement with supernova Type Ia, BAO, $H(z)$, and other cosmological data.

From the HIIG data, we find that the constraints on the non-relativistic matter density parameter \om\ are consistent with other estimates, ranging between a high of $0.300^{+0.106}_{-0.083}$ (flat XCDM) and a low of $\Omega_{\rm m_0} = 0.210^{+0.043}_{-0.092}$ (flat \pcdm).

The HIIG data constraints on $H_0$ in Table \ref{tab:1d_BFP} are consistent with the estimate of $H_0 = 71.0 \pm 2.8 ({\rm stat.}) \pm 2.1 ({\rm sys.})$ \hunit made by \cite{FernandezArenas} based on a compilation of HIIG measurements that differs from what we have used here. The \hiig\ $H_0$ constraints listed in Table \ref{tab:1d_BFP} are also consistent with other recent measurements of $H_0$, being between $0.90\sigma$ (flat XCDM) and $1.56\sigma$ (non-flat \pcdm) lower than the recent local expansion rate measurement of $H_0 = 74.03 \pm 1.42$ \hunit \citep{riess_etal_2019},\footnote{Note that other local expansion rate measurements are slightly lower with slightly larger error bars \citep{rigault_etal_2015,zhangetal2017,Dhawan,FernandezArenas,freedman_etal_2019,freedman_etal_2020,rameez_sarkar_2019}.} and between $0.78\sigma$ (non-flat \pcdm) and $1.13\sigma$ (flat XCDM) higher than the median statistics estimate of $H_0=68 \pm 2.8$ \hunit \citep{chenratmed},\footnote{This is consistent with earlier median statistics estimates \citep{gott_etal_2001,chen_etal_2003} and also with a number of recent $H_0$ measurements \citep{chen_etal_2017,DES_2018,Gomez-ValentAmendola2018,haridasu_etal_2018,planck2018b,zhang_2018,dominguez_etal_2019,martinelli_tutusaus_2019,Cuceu_2019,zeng_yan_2019,schoneberg_etal_2019,lin_ishak_2019,zhang_huang_2019}.} with our measurements ranging from a low of $H_0=70.60^{+1.68}_{-1.84}$ \hunit (non-flat \pcdm) to a high of $H_0=71.85 \pm 1.96$ \hunit (flat XCDM).

As for spatial curvature, from the marginalized 1D likelihoods in Table \ref{tab:1d_BFP}, for non-flat \lcdm, non-flat XCDM, and non-flat \pcdm, we measure $\Omega_{\rm k_0}=0.094^{+0.237}_{-0.363}$,\footnote{Since $\Omega_{\rm k_0}=1-\Omega_{\rm m_0}-\Omega_{\Lambda}$, in the non-flat \lcdm\ model analysis we replace $\Omega_{\Lambda}$ with $\Omega_{\rm k_0}$ in the MCMC chains of $\{H_0, \Omega_{\rm m_0}, \Omega_{\Lambda}\}$ to obtain new chains of $\{H_0, \Omega_{\rm m_0}, \Omega_{\rm k_0}\}$ and so measure $\Omega_{\rm k_0}$ central values and uncertainties. A similar procedure, based on $\Omega_{\Lambda}=1-\Omega_{\rm m_0}$, is used to measure $\Omega_{\Lambda}$ in the flat \lcdm\ model.} $\Omega_{\rm k_0}=0.011^{+0.457}_{-0.460}$, and $\Omega_{\rm k_0}=0.291^{+0.348}_{-0.113}$, respectively. From the marginalized likelihoods, we see that non-flat \lcdm\ and XCDM models are consistent with all three spatial geometries, while non-flat \pcdm\ favors the open case at 2.58$\sigma$. However, this seems to be a little odd, especially for non-flat \pcdm, considering their unmarginalized best-fit $\Omega_{\rm k_0}$\!'s are all negative (see Table \ref{tab:BFP}).

The fits to the HIIG data are consistent with dark energy being a cosmological constant but don't rule out dark energy dynamics. For flat (non-flat) XCDM, $w_{\rm X}=-1.180^{+0.560}_{-0.330}$ ($w_{\rm X}=-1.125^{+0.537}_{-0.321}$), which are both within 1$\sigma$ of $w_{\rm X}=-1$. For flat (non-flat) \pcdm, $2\sigma$ upper limits of $\alpha$ are $\alpha<2.784$ ($\alpha<4.590$), with the 1D likelihood functions, in both cases, peaking at $\alpha=0$.

Current HIIG data do not provide very restrictive constraints on cosmological model parameters, but when used in conjunction with other cosmological data they can help tighten the constraints.

\subsection{$H(z)$, BAO, and HIIG (HzBH) constraints}
\label{subsec:HzBH}

The HIIG constraints discussed in the previous subsection are consistent with constraints from most other cosmological data, so it is appropriate to use the HIIG data in conjunction with other data to constrain parameters. In this subsection we perform a full analysis of $H(z)$, BAO, and HIIG (HzBH) data and derive tighter constraints on cosmological parameters.

\begin{figure*}
\centering
  \subfloat[Full parameter range]{%
    \includegraphics[width=3.5in,height=3.5in]{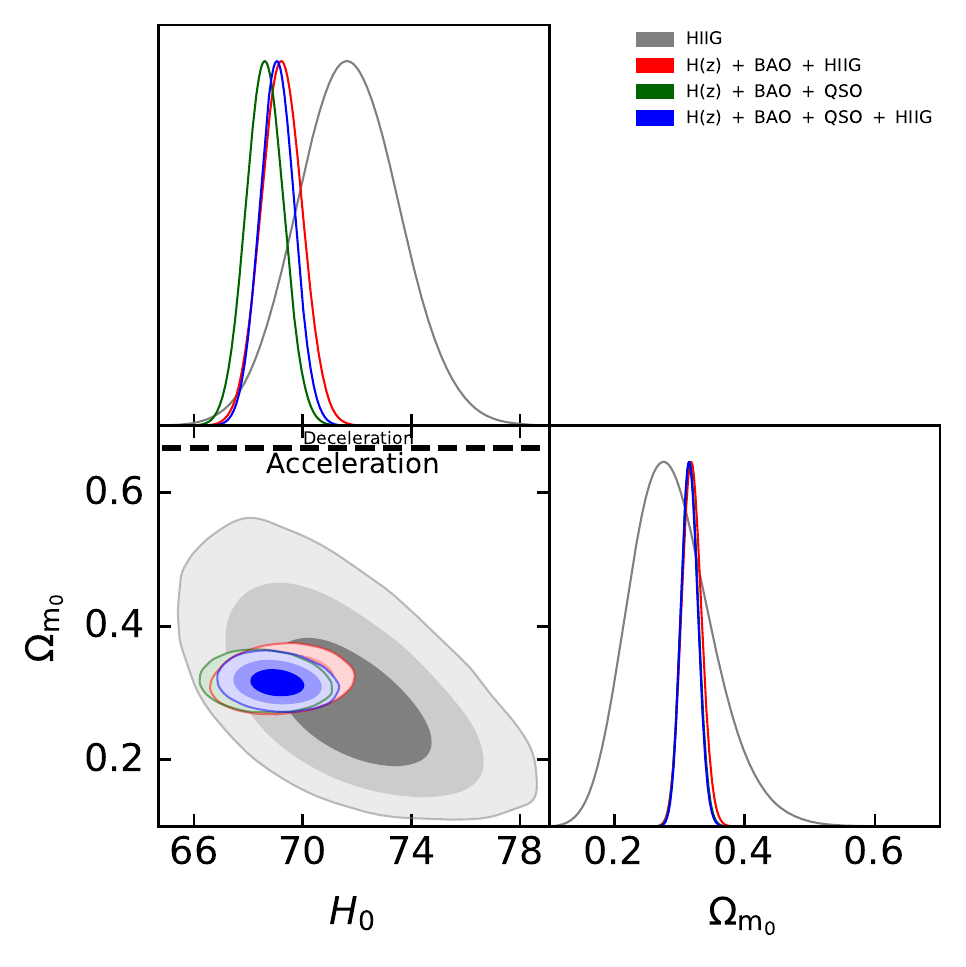}}
  \subfloat[Zoom in]{%
    \includegraphics[width=3.5in,height=3.5in]{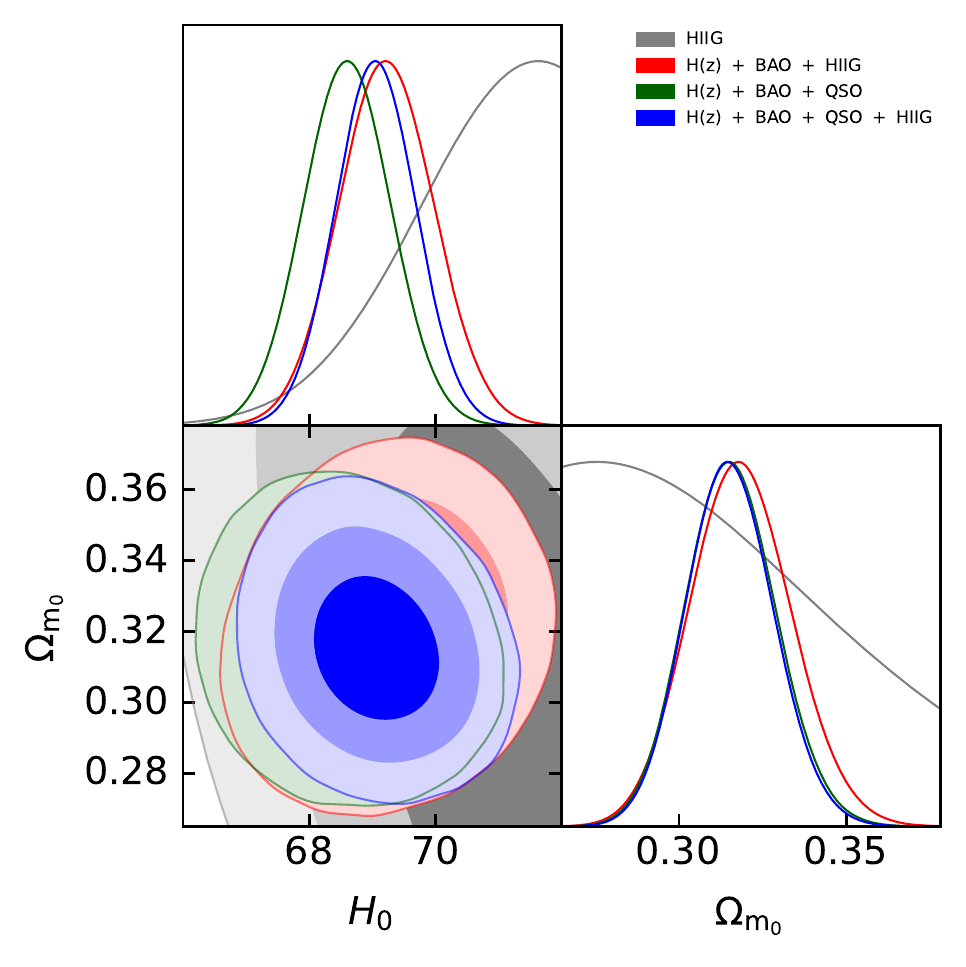}}\\
\caption{1$\sigma$, 2$\sigma$, and 3$\sigma$ confidence contours for flat \lcdm, where the right panel is the comparison zoomed in. The black dotted line is the zero-acceleration line, which divides the parameter space into regions associated with currently accelerated (below) and currently decelerated (above) cosmological expansion.}
\label{fig01}
\end{figure*}

\begin{figure*}
\centering
  \subfloat[Full parameter range]{%
    \includegraphics[width=3.5in,height=3.5in]{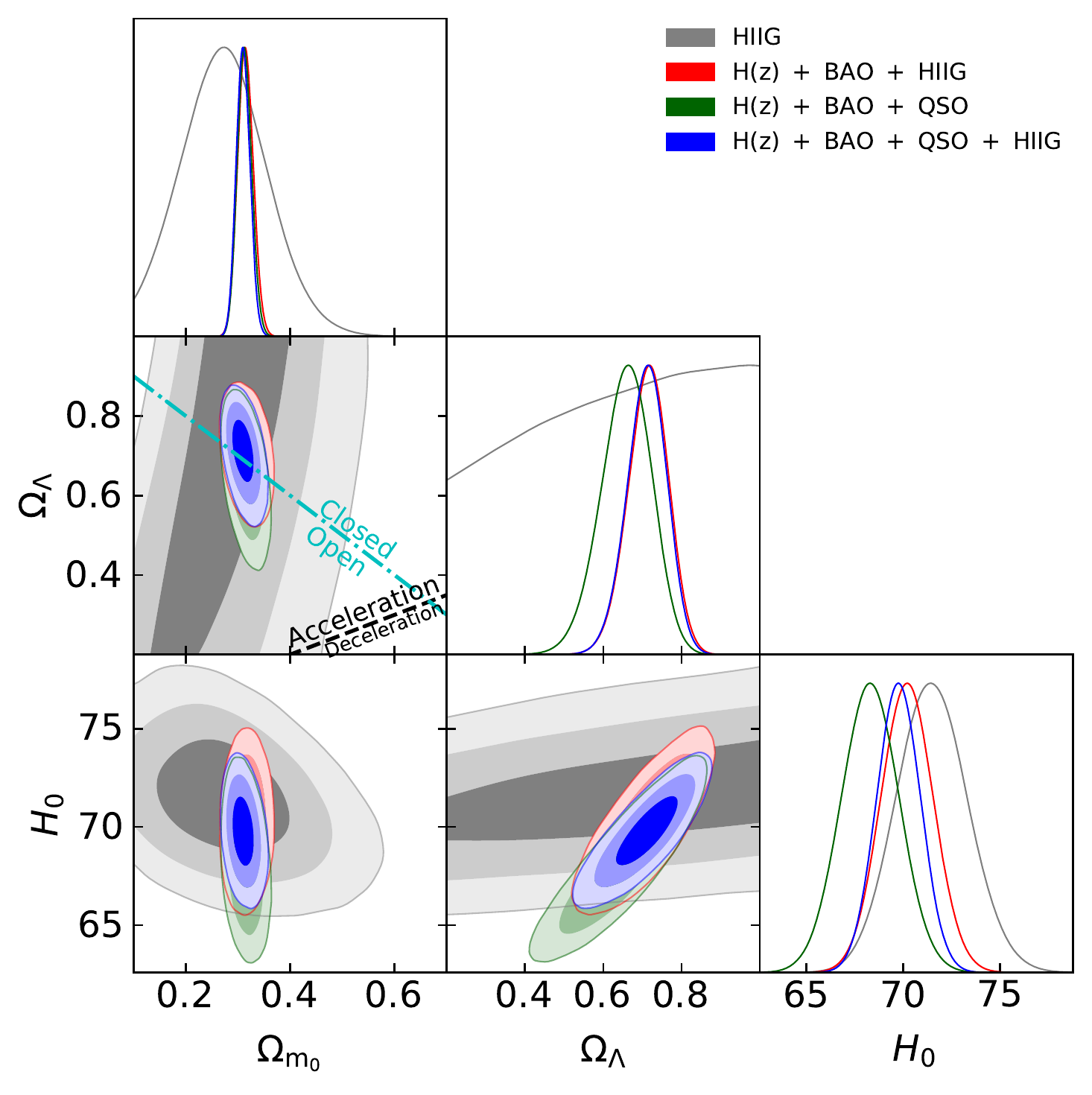}}
  \subfloat[Zoom in]{%
    \includegraphics[width=3.5in,height=3.5in]{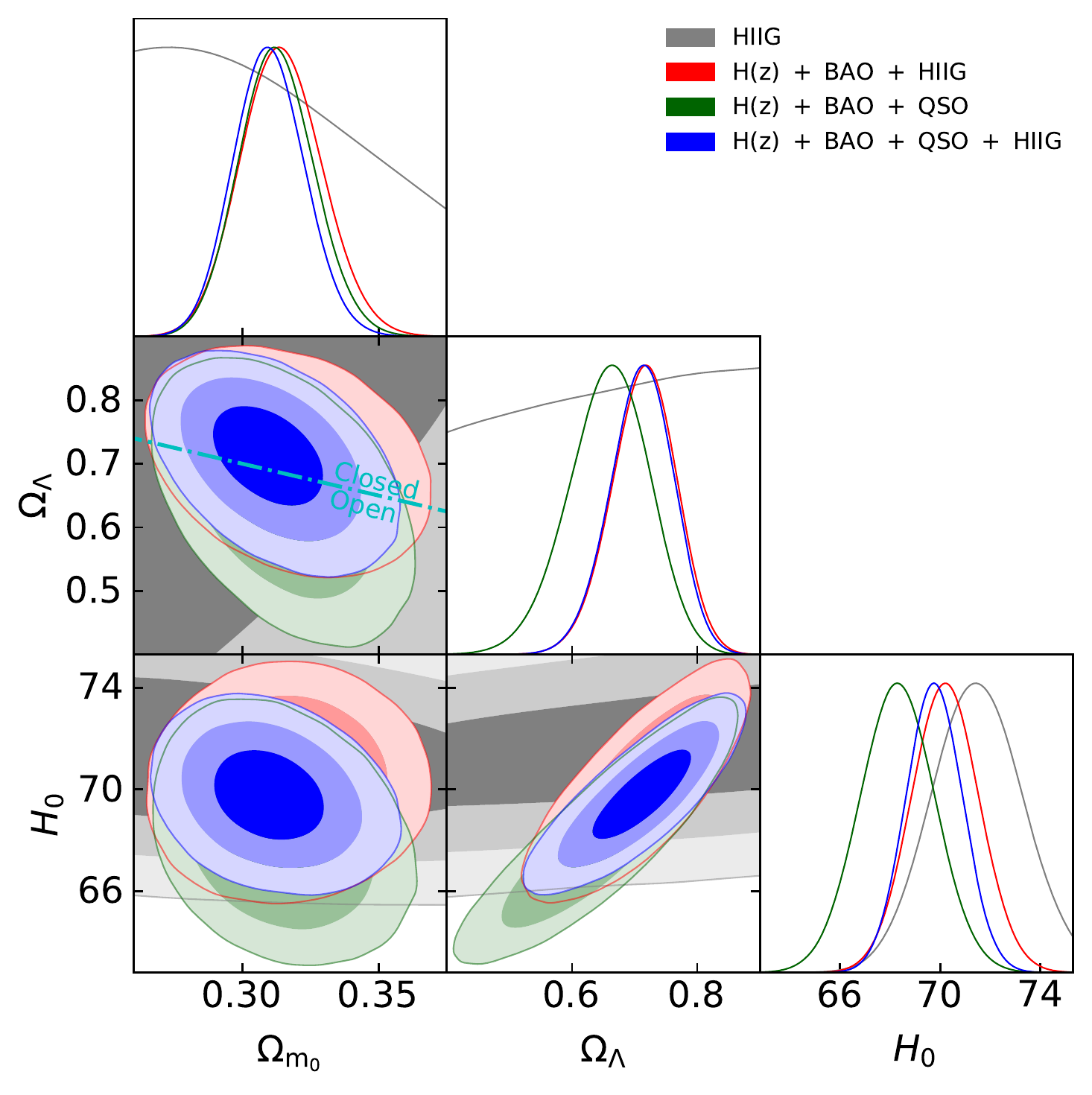}}\\
\caption{Same as Fig. \ref{fig01} but for non-flat \lcdm. The cyan dash-dot line represents the flat case, with closed spatial hypersurfaces to the upper right. The black dotted line is the zero-acceleration line, which divides the parameter space into regions associated with currently accelerated (above left) and currently decelerated (below right) cosmological expansion.}
\label{fig02}
\end{figure*}

\begin{figure*}
\centering
  \subfloat[Full parameter range]{%
    \includegraphics[width=3.5in,height=3.5in]{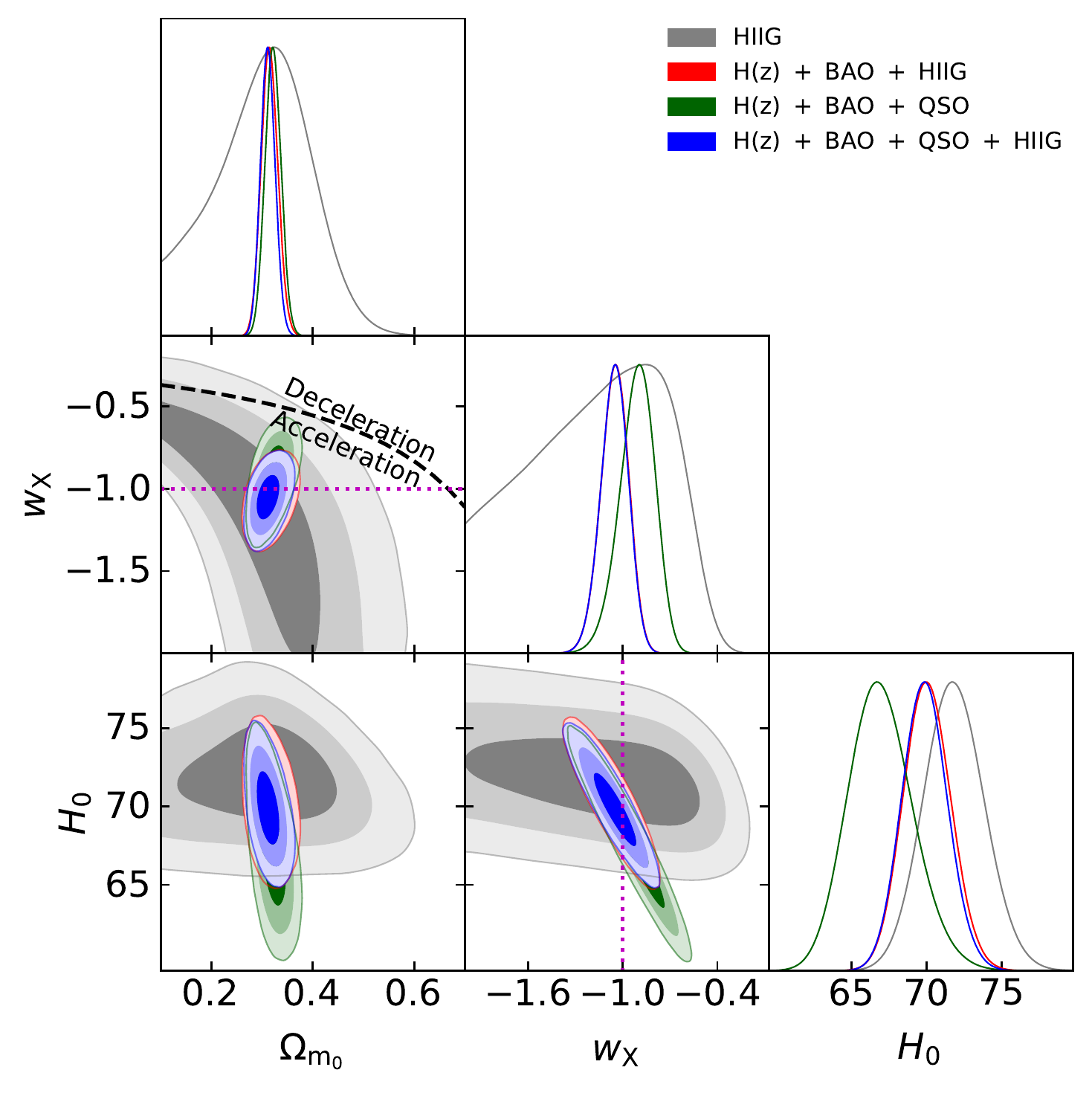}}
  \subfloat[Zoom in]{%
    \includegraphics[width=3.5in,height=3.5in]{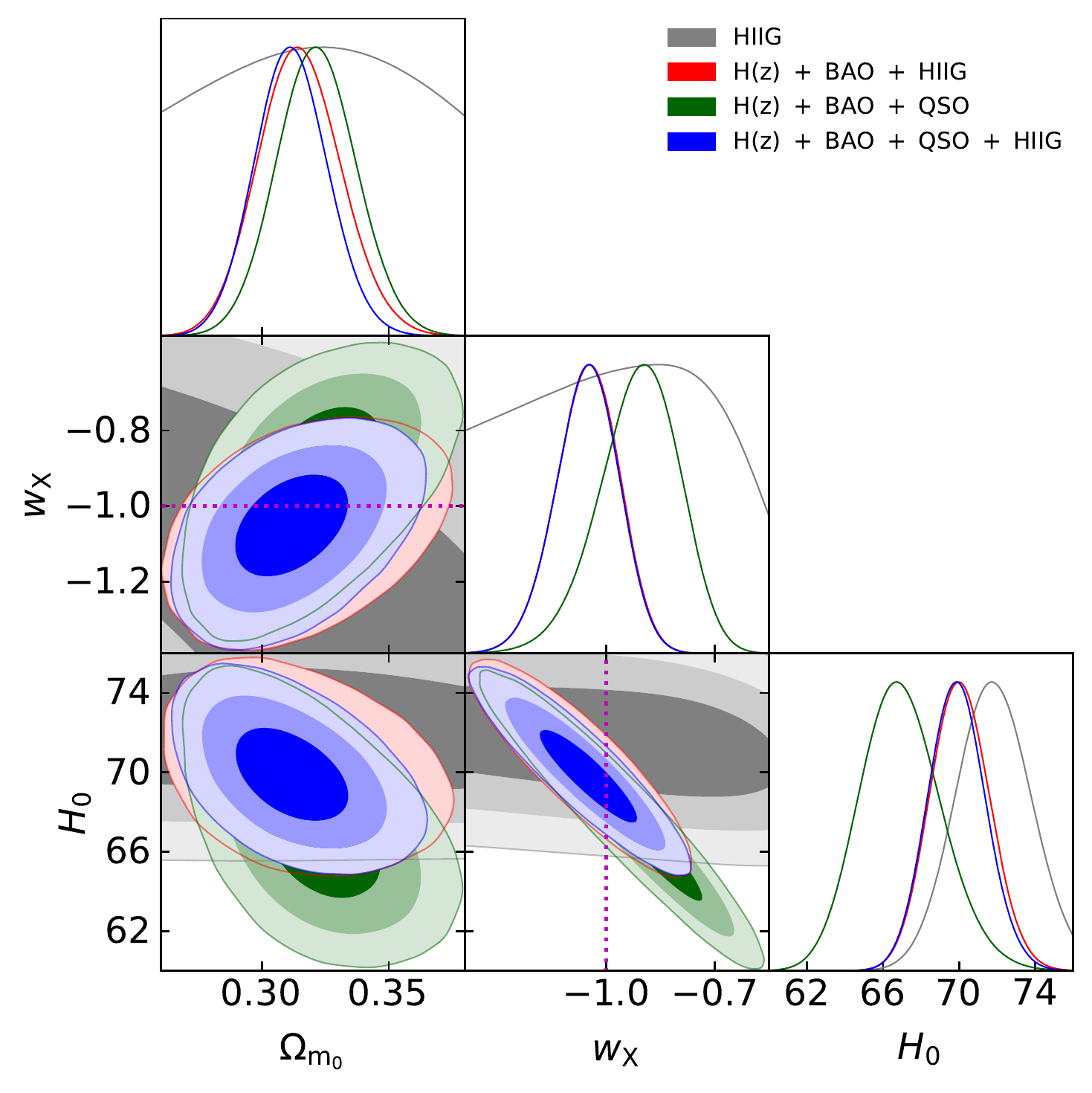}}\\
\caption{1$\sigma$, 2$\sigma$, and 3$\sigma$ confidence contours for flat XCDM. The black dotted line is the zero-acceleration line, which divides the parameter space into regions associated with currently accelerated (below left) and currently decelerated (above right) cosmological expansion. The magenta lines denote $w_{\rm X}=-1$, i.e. the flat \lcdm\ model.}
\label{fig03}
\end{figure*}

\begin{figure*}
\centering
  \subfloat[Full parameter range]{%
    \includegraphics[width=3.5in,height=3.5in]{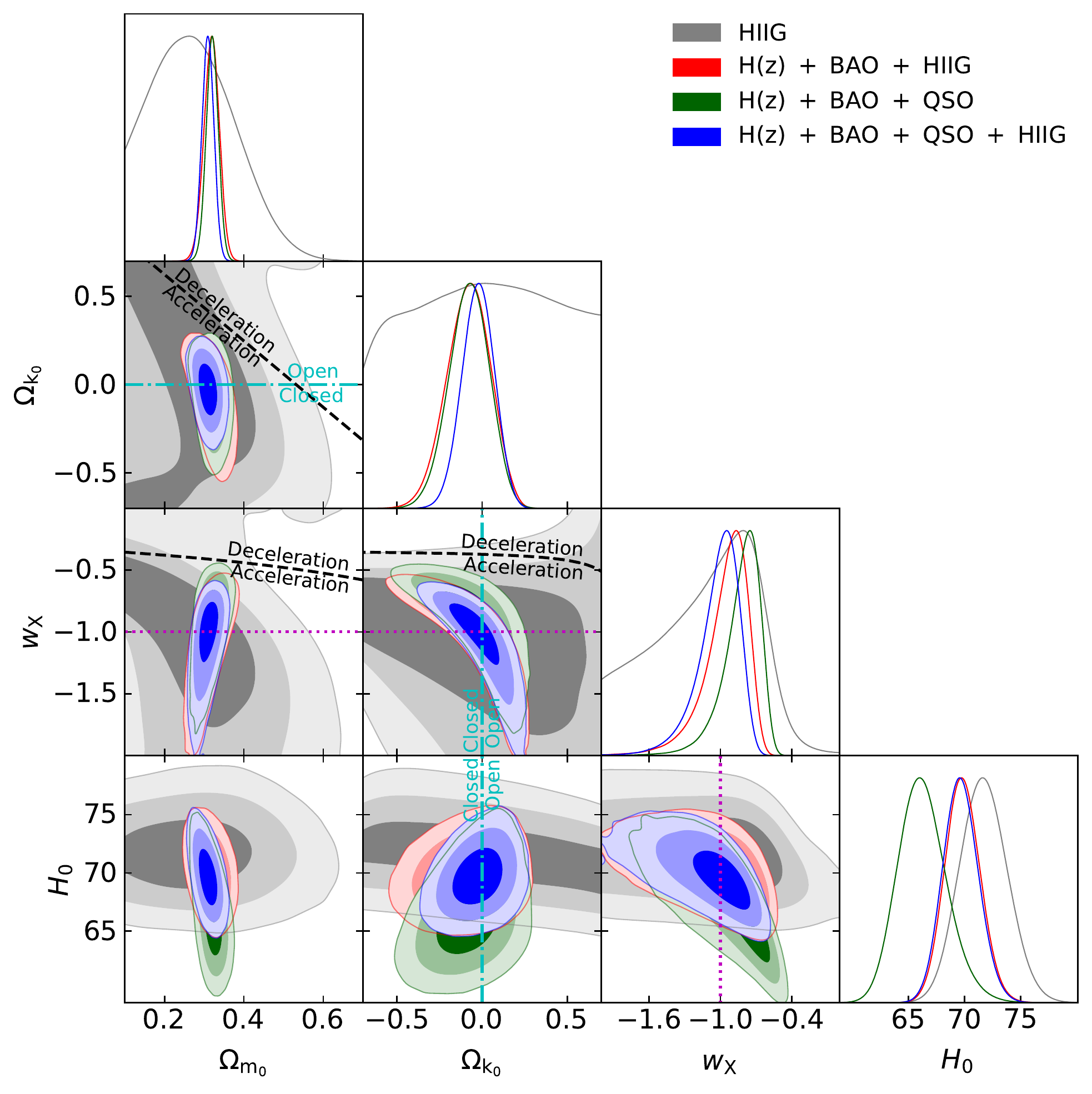}}
  \subfloat[Zoom in]{%
    \includegraphics[width=3.5in,height=3.5in]{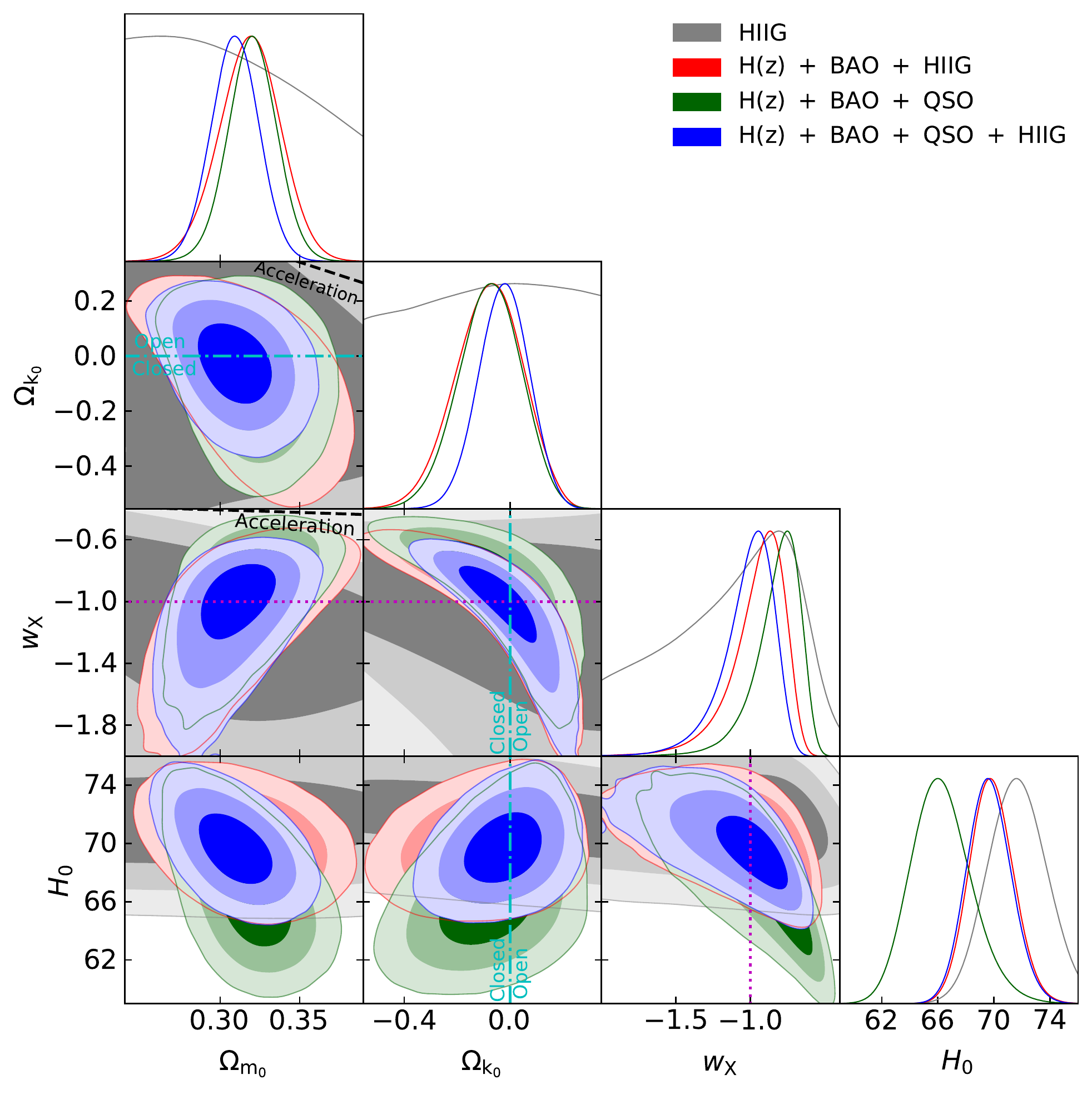}}\\
\caption{Same as Fig. \ref{fig03} but for non-flat XCDM, where the zero acceleration lines in each of the three subpanels are computed for the third cosmological parameter set to the \hiig\ data only best-fit values listed in Table \ref{tab:BFP}. Currently-accelerated cosmological expansion occurs below these lines. The cyan dash-dot lines represent the flat case, with closed spatial hypersurfaces either below or to the left. The magenta lines indicate $w_{\rm X} = -1$, i.e. the non-flat \lcdm\ model.}
\label{fig04}
\end{figure*}

\begin{figure*}
\centering
  \subfloat[Full parameter range]{%
    \includegraphics[width=3.5in,height=3.5in]{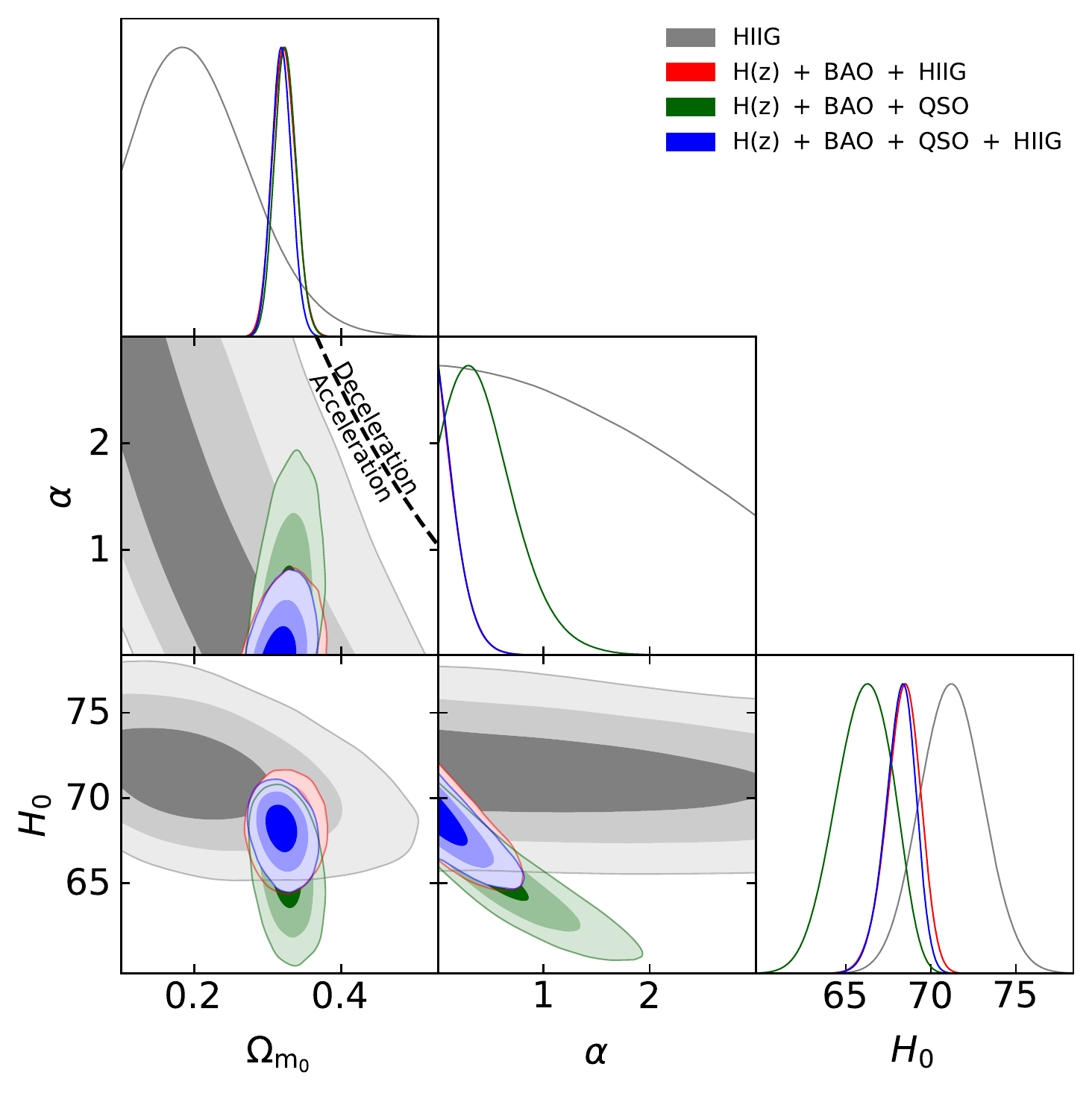}}
  \subfloat[Zoom in]{%
    \includegraphics[width=3.5in,height=3.5in]{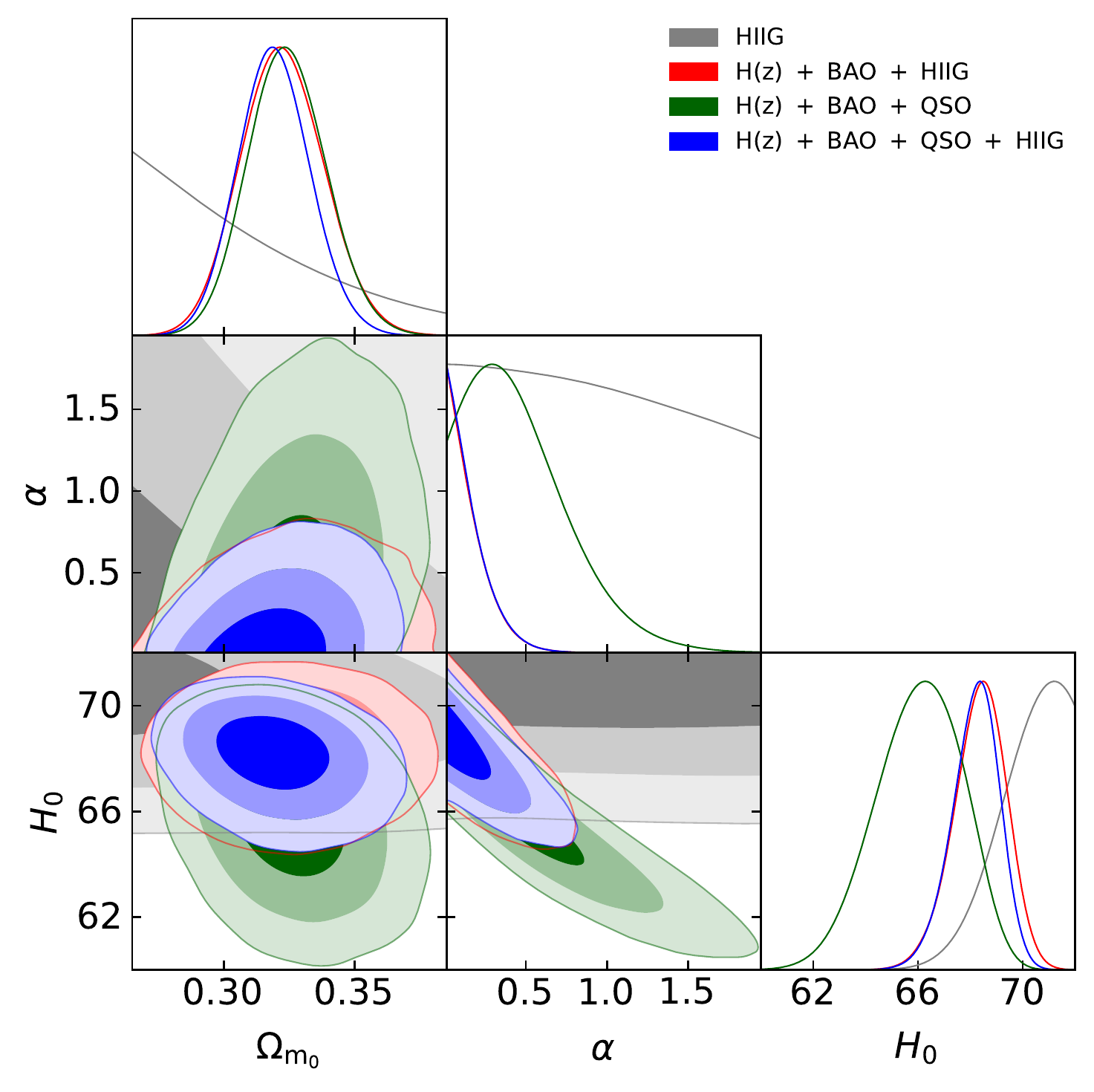}}\\
\caption{1$\sigma$, 2$\sigma$, and 3$\sigma$ confidence contours for flat $\phi$CDM. The black dotted zero-acceleration line splits the parameter space into regions of currently accelerated (below left) and currently decelerated (above right) cosmological expansion. The $\alpha = 0$ axis is the flat \lcdm\ model.}
\label{fig05}
\end{figure*}

\begin{figure*}
\centering
  \subfloat[Full parameter range]{%
    \includegraphics[width=3.5in,height=3.5in]{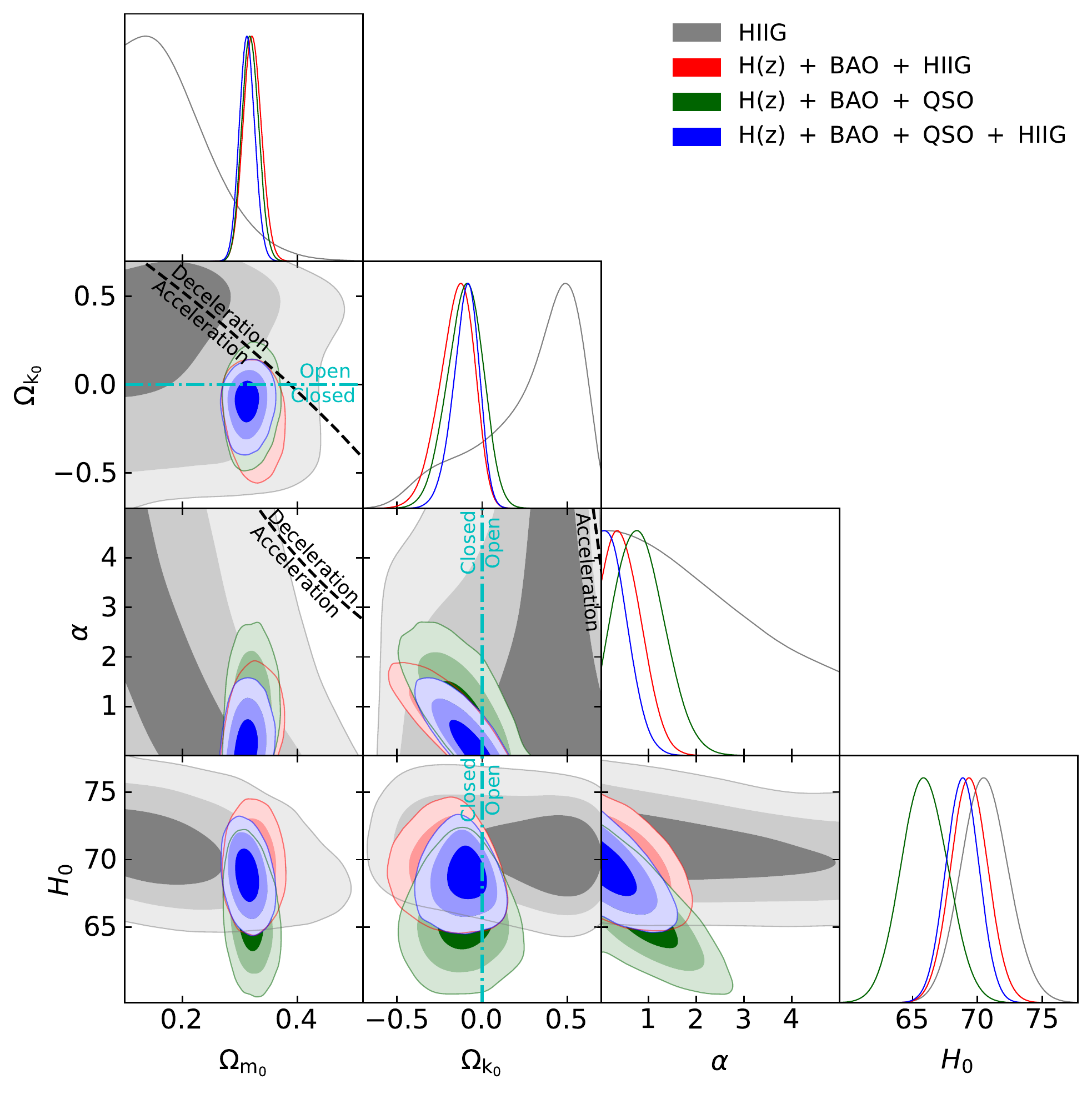}}
  \subfloat[Zoom in]{%
    \includegraphics[width=3.5in,height=3.5in]{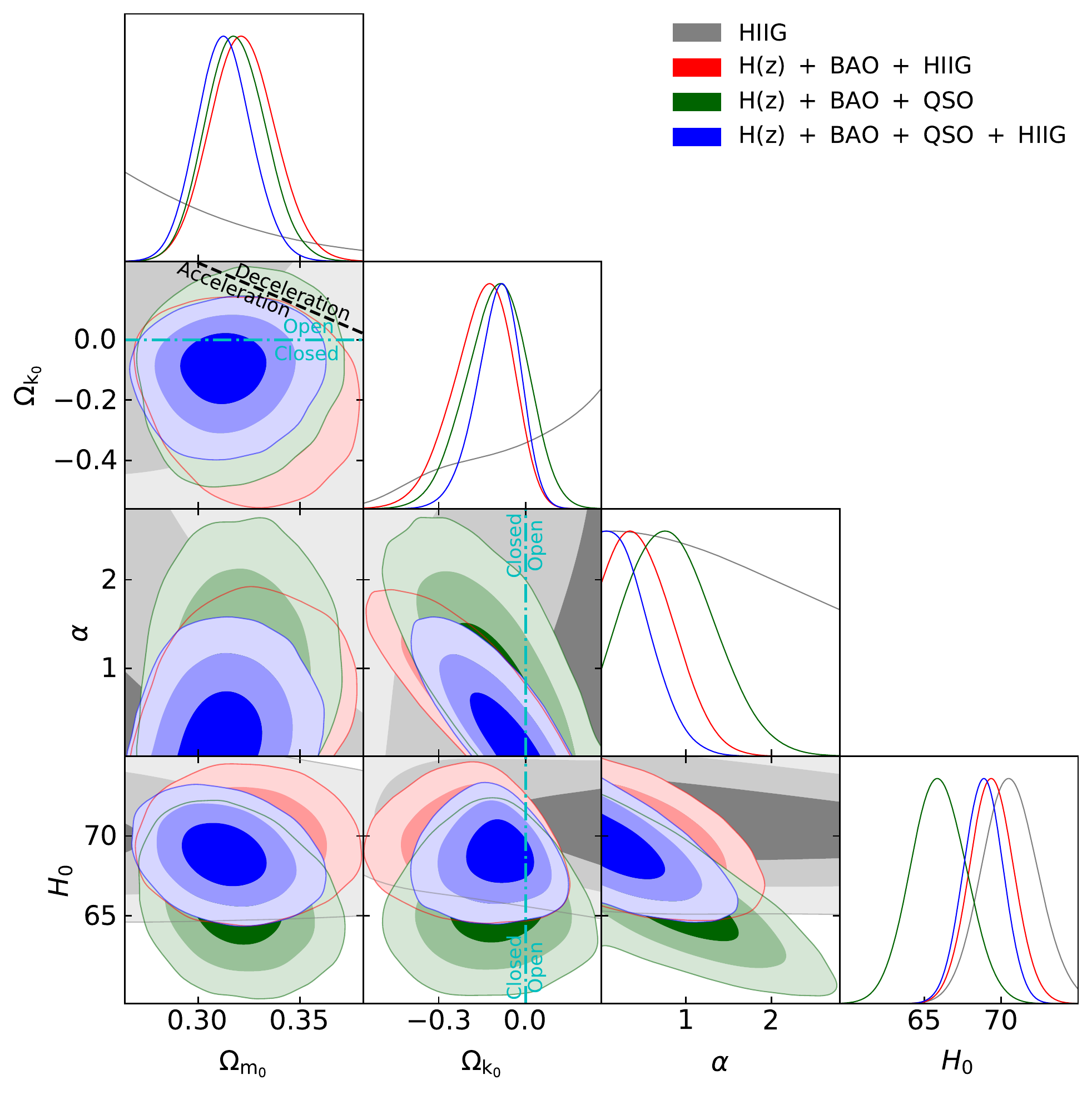}}\\
\caption{Same as Fig. \ref{fig05} but for non-flat \pcdm, where the zero-acceleration lines in each of the subpanels are computed for the third cosmological parameter set to the \hiig\ data only best-fit values listed in Table \ref{tab:BFP}. Currently-accelerating cosmological expansion occurs below these lines. The cyan dash-dot lines represent the flat case, with closed spatial geometry either below or to the left. The $\alpha = 0$ axis is the non-flat \lcdm\ model.}
\label{fig06}
\end{figure*}

The 1D probability distributions and 2D confidence regions of the cosmological parameters for the six flat and non-flat models are shown in Figs. \ref{fig01}--\ref{fig06}, in red. The best-fit results and uncertainties are listed in Tables \ref{tab:BFP} and \ref{tab:1d_BFP}.

When we fit our cosmological models to the HzBH data we find that the measured values of the matter density parameter \om\ fall within a narrower range in comparison to the HIIG only case, being between $0.314 \pm 0.015$ (non-flat \lcdm) and $0.323^{+0.014}_{-0.016}$ (flat \pcdm).

\begin{table*}
\centering
\begin{threeparttable}
\caption{Unmarginalized best-fit parameter values for all models from various combinations of data.}\label{tab:BFP}
\setlength{\tabcolsep}{2.0mm}{
\begin{tabular}{lccccccccccc}
\thickhline
 Model & Data set & $\Omega_{\mathrm m_0}$ & $\Omega_{\Lambda}$ & $\Omega_{\mathrm k_0}$ & $w_{\mathrm X}$ & $\alpha$ & $H_0$\tnote{a} & $\chi^2$ & $AIC$ & $BIC$ & $\nu$\\
\hline
Flat \lcdm & HIIG & 0.276 & 0.724 & --- & --- & --- & 71.81 & 410.75 & 414.75 & 420.81 & 151\\
 & $H(z)$ + BAO + HIIG & 0.318 & 0.682 & --- & --- & --- & 69.22 & 434.29 & 438.29 & 444.84 & 193 \\
 & $H(z)$ + BAO + QSO & 0.315 & 0.685 & --- & --- & --- & 68.61 & 372.88 & 376.88 & 383.06 & 160\\
 & $H(z)$ + BAO + QSO + HIIG & 0.315 & 0.685 & --- & --- & --- & 69.06 & 786.50 & 790.50 & 798.01 & 313\\
\hline
Non-flat \lcdm & HIIG & 0.312 & 0.998 & $-0.310$ & --- & --- & 72.35 & 410.44 & 416.44 & 425.53 & 150\\
 & $H(z)$ + BAO + HIIG & 0.313 & 0.718 & $-0.031$ & --- & --- & 70.24 & 433.38 & 439.38 & 449.19 & 192\\
 & $H(z)$ + BAO + QSO & 0.311 & 0.665 & 0.024 & --- & --- & 68.37 & 372.82 & 378.82 & 388.08 & 159\\
 & $H(z)$ + BAO + QSO + HIIG & 0.309 & 0.716 & $-0.025$ & --- & --- & 69.82 & 785.79 & 791.79 & 803.05 & 312\\
\hline
Flat XCDM & HIIG & 0.249 & --- & --- & $-0.892$ & --- & 71.65 & 410.72 & 416.72 & 425.82 & 150\\
 & $H(z)$ + BAO + HIIG & 0.314 & --- & --- & $-1.044$ & --- & 69.94 & 433.99 & 439.99 & 449.81 & 192\\
 & $H(z)$ + BAO + QSO & 0.322 & --- & --- & $-0.890$ & --- & 66.62 & 371.95 & 377.95 & 387.21 & 159\\
 & $H(z)$ + BAO + QSO + HIIG & 0.311 & --- & --- & $-1.045$ & --- & 69.80 & 786.19 & 792.19 & 803.45 & 312\\
\hline
Non-flat XCDM & HIIG & 0.104 & --- & $-0.646$ & $-0.712$ & --- & 72.61 & 407.69 & 415.69 & 427.81 & 149\\
 & $H(z)$ + BAO + HIIG & 0.322 & --- & $-0.117$ & $-0.878$ & --- & 66.67 & 432.85 & 440.85 & 453.94 & 191\\
 & $H(z)$ + BAO + QSO & 0.322 & --- & $-0.112$ & $-0.759$ & --- & 65.80 & 370.68 & 378.68 & 391.03 & 158\\
 & $H(z)$ + BAO + QSO + HIIG & 0.310 & --- & $-0.048$ & $-0.957$ & --- & 69.53 & 785.70 & 793.70 & 808.71 & 311\\
\hline
Flat $\phi$CDM & HIIG & 0.255 & --- & --- & --- & 0.261 & 71.70 & 410.70 & 416.70 & 425.80 & 150\\
 & $H(z)$ + BAO + HIIG & 0.318 & --- & --- & --- & 0.011 & 69.09 & 434.36 & 440.36 & 450.18 & 192\\
 & $H(z)$ + BAO + QSO & 0.321 & --- & --- & --- & 0.281 & 66.82 & 372.05 & 378.05 & 387.31 & 159\\
 & $H(z)$ + BAO + QSO + HIIG & 0.315 & --- & --- & --- & 0.012 & 68.95 & 786.58 & 792.58 & 803.84 & 312\\
\hline
Non-flat $\phi$CDM & HIIG & 0.114 & --- & $-0.437$ & --- & 2.680 & 72.14 & 409.91 & 417.91 & 430.03 & 149\\
 & $H(z)$ + BAO + HIIG & 0.321 & --- & $-0.132$ & --- & 0.412 & 69.69 & 432.75 & 440.75 & 453.84 & 191\\
 & $H(z)$ + BAO + QSO & 0.317 & --- & $-0.106$ & --- & 0.778 & 66.27 & 370.83 & 378.83 & 391.18 & 158\\
 & $H(z)$ + BAO + QSO + HIIG & 0.310 & --- & $-0.054$ & --- & 0.150 & 69.40 & 785.65 & 793.65 & 808.66 & 311\\
\thickhline
\end{tabular}}
\begin{tablenotes}
\item [a] \hunit.
\end{tablenotes}
\end{threeparttable}
\end{table*}

\begin{table*}
\centering
\begin{threeparttable}
\caption{One-dimensional marginalized best-fit parameter values and uncertainties ($\pm 1\sigma$ error bars or $2\sigma$ limits) for all models from various combinations of data.}\label{tab:1d_BFP}
\setlength{\tabcolsep}{1.8mm}{
\begin{tabular}{lccccccc}
\thickhline
 Model & Data set & $\Omega_{\mathrm m_0}$ & $\Omega_{\Lambda}$ & $\Omega_{\mathrm k_0}$ & $w_{\mathrm X}$ & $\alpha$ & $H_0$\tnote{a} \\
\hline
Flat \lcdm & HIIG & $0.289^{+0.053}_{-0.071}$ & --- & --- & --- & --- & $71.70\pm1.83$ \\
 & $H(z)$ + BAO + HIIG & $0.319^{+0.014}_{-0.015}$ & --- & --- & --- & --- & $69.23\pm0.74$ \\
 & $H(z)$ + BAO + QSO & $0.316^{+0.013}_{-0.014}$ & --- & --- & --- & --- & $68.60\pm0.68$ \\
 & $H(z)$ + BAO + QSO + HIIG & $0.315^{+0.013}_{-0.012}$ & --- & --- & --- & --- & $69.06^{+0.63}_{-0.62}$ \\
\hline
Non-flat \lcdm & HIIG & $0.275^{+0.081}_{-0.078}$ & $>0.501$\tnote{b} & $0.094^{+0.237}_{-0.363}$ & --- & --- & $71.50^{+1.80}_{-1.81}$ \\
 & $H(z)$ + BAO + HIIG & $0.314\pm0.015$ & $0.714^{+0.054}_{-0.049}$ & $-0.029^{+0.049}_{-0.048}$ & --- & --- & $70.21\pm1.33$ \\
 & $H(z)$ + BAO + QSO & $0.313^{+0.013}_{-0.015}$ & $0.658^{+0.069}_{-0.060}$ & $0.029^{+0.056}_{-0.063}$ & --- & --- & $68.29\pm1.47$ \\
 & $H(z)$ + BAO + QSO + HIIG & $0.310\pm0.013$ & $0.711^{+0.053}_{-0.048}$ & $-0.021^{+0.044}_{-0.048}$ & --- & --- & $69.76^{+1.12}_{-1.11}$ \\
\hline
Flat XCDM & HIIG & $0.300^{+0.106}_{-0.083}$ & --- & --- & $-1.180^{+0.560}_{-0.330}$ & --- & $71.85\pm1.96$ \\
 & $H(z)$ + BAO + HIIG & $0.315^{+0.016}_{-0.017}$ & --- & --- & $-1.052^{+0.092}_{-0.082}$ & --- & $70.05\pm1.54$ \\
 & $H(z)$ + BAO + QSO & $0.322^{+0.015}_{-0.016}$ & --- & --- & $-0.911^{+0.122}_{-0.098}$ & --- & $66.98^{+1.95}_{-2.30}$ \\
 & $H(z)$ + BAO + QSO + HIIG & $0.312\pm0.014$ & --- & --- & $-1.053^{+0.091}_{-0.082}$ & --- & $69.90\pm1.48$ \\
\hline
Non-flat XCDM & HIIG & $0.275^{+0.084}_{-0.125}$ & --- & $0.011^{+0.457}_{-0.460}$ & $-1.125^{+0.537}_{-0.321}$ & --- & $71.71^{+2.07}_{-2.08}$ \\
 & $H(z)$ + BAO + HIIG & $0.318\pm0.019$ & --- & $-0.082^{+0.135}_{-0.119}$ & $-0.958^{+0.219}_{-0.098}$ & --- & $69.83^{+1.50}_{-1.62}$ \\
 & $H(z)$ + BAO + QSO & $0.320\pm0.015$ & --- & $-0.078^{+0.124}_{-0.112}$ & $-0.826^{+0.185}_{-0.088}$ & --- & $66.29^{+1.90}_{-2.35}$ \\
 & $H(z)$ + BAO + QSO + HIIG & $0.309^{+0.015}_{-0.014}$ & --- & $-0.025\pm0.092$ & $-1.022^{+0.208}_{-0.104}$ & --- & $69.68^{+1.49}_{-1.64}$ \\
\hline
Flat $\phi$CDM & HIIG & $0.210^{+0.043}_{-0.092}$ & --- & --- & --- & $<2.784$ & $71.23^{+1.79}_{-1.80}$ \\
 & $H(z)$ + BAO + HIIG & $0.323^{+0.014}_{-0.016}$ & --- & --- & --- & $<0.411$ & $68.36^{+1.05}_{-0.86}$ \\
 & $H(z)$ + BAO + QSO & $0.324^{+0.014}_{-0.015}$ & --- & --- & --- & $0.460^{+0.116}_{-0.440}$ & $66.03^{+1.79}_{-1.42}$ \\
 & $H(z)$ + BAO + QSO + HIIG & $0.319\pm0.013$ & --- & --- & --- & $<0.411$ & $68.18^{+0.97}_{-0.75}$\\
\hline
Non-flat $\phi$CDM & HIIG & $<0.321$ & --- & $0.291^{+0.348}_{-0.113}$ & --- & $<4.590$ & $70.60^{+1.68}_{-1.84}$ \\
 & $H(z)$ + BAO + HIIG & $0.322^{+0.015}_{-0.016}$ & --- & $-0.153^{+0.114}_{-0.079}$ & --- & $0.538^{+0.151}_{-0.519}$ & $69.39\pm1.37$ \\
 & $H(z)$ + BAO + QSO & $0.319^{+0.013}_{-0.015}$ & --- & $-0.103^{+0.111}_{-0.091}$ & --- & $0.854^{+0.379}_{-0.594}$ & $65.94^{+1.75}_{-1.73}$ \\
 & $H(z)$ + BAO + QSO + HIIG & $0.313^{+0.012}_{-0.014}$ & --- & $-0.098^{+0.082}_{-0.061}$ & --- & $<0.926$ & $68.83\pm1.23$ \\
\thickhline
\end{tabular}}
\begin{tablenotes}
\item [a] \hunit.
\item [b] This is the 1$\sigma$ lower limit. The $2\sigma$ lower limit is set by the prior, and is not shown here.
\end{tablenotes}
\end{threeparttable}
\end{table*}

Similarly, the measured values of $H_0$ also fall within a narrower range when our models are fit to the HzBH data combination (and are in better agreement with the median statistics estimate of $H_0$ from \citealp{chenratmed} than with the local measurement carried out by \citealp{riess_etal_2019}; this is because the $H(z)$ and BAO data favor a lower $H_0$ value) being between $H_0=68.36^{+1.05}_{-0.86}$ \hunit (flat \pcdm) and $70.21 \pm 1.33$ \hunit (non-flat \lcdm). We assume that the tension between early- and late-Universe measurements of $H_0$ is not a major issue here, because the 2D and 1D contours in Fig. \ref{fig01} overlap, and so we compute a combined $H_0$ value (but if one is concerned about the early- vs late-Universe $H_0$ tension then one should not compare our combined-data $H_0$\!'s here, and in Secs. \ref{subsec:HzBQ} and \ref{subsec:HzBQH}, directly to the measurements of \citealp{riess_etal_2019} or of \citealp{planck2018b}).

In contrast to the HIIG only cases, when fit to the HzBH data combination the non-flat models mildly favor closed spatial hypersurfaces. This is because the $H(z)$ and BAO data mildly favor closed spatial hypersurfaces; see, e.g. \cite{park_ratra_2019b} and \cite{Ryan_2}. For non-flat \lcdm, non-flat XCDM, and non-flat \pcdm, we find $\Omega_{\rm k_0}=-0.029^{+0.049}_{-0.048}$, $\Omega_{\rm k_0}=-0.082^{+0.135}_{-0.119}$, and $\Omega_{\rm k_0}=-0.153^{+0.114}_{-0.079}$, respectively, with the non-flat \pcdm\ model favoring closed spatial hypersurfaces at 1.34$\sigma$.

The fit to the HzBH data combination produces weaker evidence for dark energy dynamics (in comparison to the HIIG only case) with tighter error bars on the measured values of $w_{\rm X}$ and $\alpha$. For flat (non-flat) XCDM, $w_{\rm X}=-1.052^{+0.092}_{-0.082}$ ($w_{\rm X}=-0.958^{+0.219}_{-0.098}$), with $w_{\rm X}=-1$ still being within the 1$\sigma$ range. For flat (non-flat) \pcdm, $\alpha<0.411$ ($\alpha=0.538^{+0.151}_{-0.519}$), where the former is peaked at $\alpha=0$ but for the latter, $\alpha=0$ is just out of the 1$\sigma$ range.

\subsection{$H(z)$, BAO, and QSO (HzBQ) constraints}
\label{subsec:HzBQ}

The $H(z)$, BAO, and QSO (HzBQ) data combination has previously been studied \citep{Ryan_2}. Relative to that analysis, we use an updated BAO data compilation, a more accurate formula for $r_s$, and the MCMC formalism (instead of the grid-based $\chi^2$ approach); consequently the parameter constraints derived here slightly differ from those of \cite{Ryan_2}.

The 1D probability distributions and 2D confidence regions of the cosmological parameters for all models are presented in Figs. \ref{fig01}--\ref{fig06}, in green. The corresponding best-fit results and uncertainties are listed in Tables \ref{tab:BFP} and \ref{tab:1d_BFP}.

The measured values of \om\ here fall within a similar range to the range quoted in the last subsection, being between $0.313^{+0.013}_{-0.015}$ (non-flat \lcdm) and $0.324^{+0.014}_{-0.015}$ (flat \pcdm). This range is larger than, but still consistent with, the range of \om\ reported in \cite{Ryan_2}, where the same models are fit to the HzBQ data combination.

The $H_0$ measurements in this case fall within a broader range than in the HzBH case, being between $65.94^{+1.75}_{-1.73}$ \hunit (non-flat \pcdm) and $68.60 \pm 0.68$ \hunit (flat \lcdm). In addition, they are lower than the corresponding measurements in the HzBH cases, and are in better agreement with the median statistics \citep{chenratmed} estimate of $H_0$ than with what is measured from the local expansion rate \citep{riess_etal_2019}. Compared with \cite{Ryan_2}, the central values are lower except for the non-flat XCDM model.

For non-flat \lcdm, non-flat XCDM, and non-flat \pcdm, we measure $\Omega_{\rm k_0}=0.029^{+0.056}_{-0.063}$, $\Omega_{\rm k_0}=-0.078^{+0.124}_{-0.112}$, and $\Omega_{\rm k_0}=-0.103^{+0.111}_{-0.091}$, respectively. These results are consistent with their unmarginalized best-fits (see Table \ref{tab:BFP}), where the best-fit to the non-flat \lcdm\ model favors open spatial hypersurfaces, and the best-fits to the non-flat XCDM parametrization and the non-flat \pcdm\ model both favor closed spatial hypersurfaces. Note that the central values are larger than those of \cite{Ryan_2}, especially for non-flat \lcdm\ (positive instead of negative). In all three models the constraints are consistent with flat spatial hyperfurfaces.

The fit to the HzBQ data combination provides slightly stronger evidence for dark energy dynamics than does the fit to the HzBH data combination. For flat (non-flat) XCDM, $w_{\rm X}=-0.911^{+0.122}_{-0.098}$ ($w_{\rm X}=-0.826^{+0.185}_{-0.088}$), with the former barely within 1$\sigma$ of $w_{\rm X}=-1$ and the latter almost 2$\sigma$ away from $w_{\rm X}=-1$. For flat (non-flat) \pcdm, $\alpha=0.460^{+0.116}_{-0.440}$ ($\alpha=0.854^{+0.379}_{-0.594}$), with the former 1.05$\sigma$ and the latter 1.44$\sigma$ away from the $\alpha=0$ cosmological constant. In comparison with \cite{Ryan_2}, central values of $w_{\rm X}$ are larger and smaller for flat and non-flat XCDM models, respectively, and that of $\alpha$ are larger for both flat and non-flat \pcdm\ models.

\subsection{$H(z)$, BAO, QSO, and HIIG (HzBQH) constraints}
\label{subsec:HzBQH}

Comparing the results of the previous two subsections, we see that when used in conjunction with $H(z)$ and BAO data, the QSO data result in tighter constraints on $\Omega_{\rm m_0}$, $\Omega_{\rm k_0}$ (in non-flat XCDM), $w_{\rm X}$ (in non-flat XCDM), and $H_0$ (in flat \lcdm), while the HIIG data result in tighter constraints on $H_0$ (except for flat \lcdm), $\Omega_{\Lambda}$, $\Omega_{\rm k_0}$(in non-flat \lcdm\ and \pcdm), $w_{\rm X}$ (in flat XCDM), and $\alpha$. Consequently, it is useful to derive constraints from an analysis of the combined $H(z)$, BAO, QSO, and HIIG (HzBQH) data. We present the results of such an analysis in this subsection.

In Figs. \ref{fig01}--\ref{fig06}, we present the 1D probability distributions and 2D confidence constraints for the HzBQH cases in blue. Tables \ref{tab:BFP} and \ref{tab:1d_BFP} list the best-fit results and uncertainties.

It is interesting that the best-fit values of $\Omega_{\rm m_0}$ in this case are lower compared with both the HzBQ and the HzBH results, being between $0.309^{+0.015}_{-0.014}$ (non-flat XCDM) and $0.319 \pm 0.013$ (flat \pcdm). The best-fit values of $H_0$ are higher than the HzBQ cases and have central values that are closer to those of the HzBH cases, but are still in better agreement with the lower median statistics estimate of $H_0$ \citep{chenratmed} than the higher local expansion rate measurement of $H_0$ \citep{riess_etal_2019}, being between $68.18^{+0.97}_{-0.75}$ \hunit (flat \pcdm) and $69.90 \pm 1.48$ \hunit (flat XCDM). 

For non-flat \lcdm, non-flat XCDM, and non-flat \pcdm, we measure $\Omega_{\rm k_0}=-0.021^{+0.044}_{-0.048}$, $\Omega_{\rm k_0}=-0.025 \pm 0.092$, and $\Omega_{\rm k_0}=-0.098^{+0.082}_{-0.061}$, respectively. For non-flat \lcdm\ and XCDM, the measured values of the curvature energy density parameter are within 0.48$\sigma$ and 0.27$\sigma$ of $\Omega_{\rm k_0} = 0$, respectively, while the non-flat \pcdm\ model favors a closed geometry with an $\Omega_{\rm k_0}$ that is 1.20$\sigma$ away from zero.

There is not much evidence in support of dark energy dynamics in the HzBQH case, with $\Lambda$ consistent with this data combination. For flat (non-flat) XCDM, $w_{\rm X}=-1.053^{+0.091}_{-0.082}$ ($w_{\rm X}=-1.022^{+0.208}_{-0.104}$). For flat (non-flat) \pcdm, the $2\sigma$ upper limits are $\alpha<0.411$ ($\alpha<0.926$), which indicates that $\alpha = 0$ or $\Lambda$ is consistent with these data.

\subsection{Model comparison}
\label{sec:comparison}

From Table \ref{tab:cab}, we see that the reduced $\chi^2$ for all models is relatively large (being between 2.25 and 2.75). This could probably be attributed to underestimated systematic uncertainties in the HIIG data.\footnote{Underestimated systematic uncertainties might also explain the large reduced $\chi^2$ of QSO data \citep{Ryan_2}.} This is suggested by \cite{G-M_2019}, who also found relatively large values of $\chi^2/\nu$ in their cosmological model fits to the HIIG data (though not as large as ours, because they compute a different $\chi^2$, as explained in footnote \ref{fn5} in Sec. \ref{sec:analysis}). They note that an additional systematic uncertainty of $\sim0.22$ could bring their $\chi^2/\nu$ down to $\sim1$. As mentioned previously, we do not account for HIIG systematic uncertainties in our analysis.

\begin{table*}
\centering
\caption{$\Delta \chi^2$, $\Delta AIC$, $\Delta BIC$, and $\chi^2_{\mathrm{min}}/\nu$ values.}\label{tab:cab}
\setlength{\tabcolsep}{2.0mm}{
\begin{tabular}{lccccccc}
\thickhline
 Quantity & Data set & Flat \lcdm & Non-flat \lcdm & Flat XCDM & Non-flat XCDM & Flat \pcdm & Non-flat \pcdm\\
\hline
 & HIIG & 3.06 & 2.75 & 3.03 & 0.00 & 3.01 & 2.22\\
$\Delta \chi^2$ & $H(z)$ + BAO + HIIG & 1.54 & 0.63 & 1.24 & 0.10 & 1.61 & 0.00 \\
 & $H(z)$ + BAO + QSO & 2.20 & 2.14 & 1.27 & 0.00 & 1.37 & 0.15\\
 & $H(z)$ + BAO + QSO + HIIG & 0.85 & 0.14 & 0.54 & 0.05 & 0.93 & 0.00\\
 \hline
 & HIIG & 0.00 & 1.69 & 1.97 & 0.94 & 1.95 & 3.16\\
$\Delta AIC$ & $H(z)$ + BAO + HIIG & 0.00 & 1.09 & 1.70 & 2.56 & 2.07 & 2.46\\
 & $H(z)$ + BAO + QSO & 0.00 & 1.94 & 1.07 & 1.80 & 1.17 & 1.95\\
 & $H(z)$ + BAO + QSO + HIIG & 0.00 & 1.29 & 1.69 & 3.20 & 2.08 & 3.15\\
 \hline
 & HIIG & 0.00 & 4.72 & 5.01 & 7.00 & 4.99 & 9.22\\
$\Delta BIC$ & $H(z)$ + BAO + HIIG & 0.00 & 4.35 & 4.97 & 9.10 & 5.34 & 9.00\\
 & $H(z)$ + BAO + QSO & 0.00 & 5.02 & 4.15 & 7.97 & 4.25 & 8.12\\
 & $H(z)$ + BAO + QSO + HIIG & 0.00 & 5.04 & 5.44 & 10.70 & 5.83 & 10.65\\
 \hline
 & HIIG & 2.72 & 2.74 & 2.74 & 2.74 & 2.74 & 2.75\\
$\chi^2_{\mathrm{min}}/\nu$ & $H(z)$ + BAO + HIIG & 2.25 & 2.26 & 2.26 & 2.27 & 2.26 & 2.27\\
 & $H(z)$ + BAO + QSO & 2.33 & 2.34 & 2.34 & 2.35 & 2.34 & 2.35\\
 & $H(z)$ + BAO + QSO + HIIG & 2.51 & 2.52 & 2.52 & 2.53 & 2.52 & 2.53\\
\thickhline
\end{tabular}}
\end{table*}

One thing that is clear, regardless of the absolute size of \hiig\ or QSO systematics (and ignoring the large values of $\chi^2/\nu$), is that the flat \lcdm\ model remains the most favored model among the six models we studied, based on the $AIC$ and $BIC$ criteria (see Table \ref{tab:cab}).\footnote{Note that based on the $\Delta \chi^2$ results of Table \ref{tab:cab} non-flat XCDM has the minimum $\chi^2$ in the HIIG and HzBQ cases, whereas non-flat \pcdm\ has the minimum $\chi^2$ for the HzBH and HzBQH cases. The $\Delta \chi^2$ values do not, however, penalize a model for having more parameters.} In Table \ref{tab:cab} we define $\Delta \chi^2$, $\Delta AIC$, and $\Delta BIC$, respectively, as the differences between the values of the $\chi^2$, $AIC$, and $BIC$ associated with a given model and their corresponding minimum values among all models.

From the HIIG results for $\Delta AIC$ and $\Delta BIC$ listed in Table \ref{tab:cab}, we see that the evidence against non-flat \lcdm, flat XCDM, and flat \pcdm\ is weak (according to $\Delta AIC$) and positive (according to $\Delta BIC$) where, among these three models, the flat XCDM model is the least favored. The evidence against the non-flat XCDM model is weak regarding $\Delta AIC$ but strong based on $\Delta BIC$, while the evidence against non-flat \pcdm\ in this case is positive ($\Delta AIC$) and strong ($\Delta BIC$), respectively, with it being the least favored model overall.

Largely similar conclusions result from $\Delta AIC$ and $\Delta BIC$ values for the HIIG and HzBQ data. The exception is that the HzBQ $\Delta AIC$ value gives only weak evidence against non-flat \pcdm, instead of the positive evidence against it from the HIIG $\Delta AIC$ value.

The HzBH and HzBQH values of $\Delta AIC$ and $\Delta BIC$ result in the following conclusions:

1) the evidence against both non-flat \lcdm\ and flat XCDM is weak (HzBH) and positive (HzBQH) for $\Delta AIC$ and $\Delta BIC$;

2) the evidence against flat \pcdm\ is positive; 

3) non-flat XCDM is the least favored model with non-flat \pcdm\ doing almost as badly. $\Delta AIC$ gives positive evidence against non-flat XCDM and non-flat \pcdm, while $\Delta BIC$ strongly disfavors (HzBH) and very strongly disfavors (HzBQH) both of these nonflat models.

\section{Conclusion}
\label{sec:conclusion}
In this paper, we have constrained cosmological parameters in six flat and non-flat cosmological models by analyzing a total of 315 observations, comprising 31 $H(z)$, 11 BAO, 120 QSO, and 153 HIIG measurements. The QSO angular size and HIIG apparent magnitude measurements are particularly noteworthy, as they reach to $z\sim2.7$ and $z\sim2.4$ respectively (somewhat beyond the highest $z\sim2.3$ reached by BAO data) and into a much less studied area of redshift space. While the current \hiig\ and QSO data do not provide very restrictive constraints, they do tighten the limits when they are used in conjunction with BAO + $H(z)$ data.

By measuring cosmological parameters in a variety of cosmological models, we are able to draw some relatively model-independent conclusions (i.e. conclusions that do not differ significantly between the different models). Specifically, for the full data set (i.e the HzBQH data), we find quite restrictive constraints on \om, a reasonable summary perhaps being $\Omega_{\rm m_0}=0.310 \pm 0.013$, in good agreement with many other recent estimates. $H_0$ is also fairly tightly constrained, with a reasonable summary perhaps being $H_0=69.5 \pm 1.2$ \hunit, which is in better agreement with the results of \cite{chenratmed} and \cite{planck2018b} than that of \cite{riess_etal_2019}. The HzBQH measurements are consistent with the standard spatially-flat \lcdm\ model, but do not strongly rule out mild dark energy dynamics or a little spatial curvature energy density. More and better-quality \hiig, QSO, and other data at $z \sim 2$--4 will significantly help to test these extensions.

\section*{Acknowledgements}

We thank Ana Luisa Gonz\'{a}lez-Mor\'{a}n and Ricardo Ch\'{a}vez for useful information and discussions related to the HIIG data, and Javier de Cruz P\'{e}rez and Chan-Gyung Park for useful discussions on the BAO data. Additionally, we thank Adam Riess for his comments on an early version of this paper, and the anonymous referee for useful suggestions. This work was partially funded by Department of Energy grants DE-SC0019038 and DE-SC0011840. Some of the computing for this project was performed on the Beocat Research Cluster at Kansas State University, which is funded in part by NSF grants CNS-1006860, EPS-1006860, EPS-0919443, ACI-1440548, CHE-1726332, and NIH P20GM113109.

\section*{Data availability}

The HIIG data underlying this article were provided to us by the authors of \cite{G-M_2019}. These data will be shared on request to the corresponding author with the permission of the authors of \cite{G-M_2019}.




\bibliographystyle{mnras}
\bibliography{mybibfile} 








\bsp	
\label{lastpage}
\end{document}